\documentclass[twocolumn,showpacs,amsmath,amssymb]{revtex4}
\usepackage{graphicx}
\usepackage{dcolumn}
\usepackage{bm}
\usepackage{relsize}
\usepackage{amsmath}

\newcommand{\beq}{\begin{equation}}
\newcommand{\eeq}{\end{equation}}
\newcommand{\ud}{\uparrow(\downarrow)}

\begin{document}

\title{Spin transport theory in ferromagnet/semiconductor systems with non-collinear magnetization configurations}
\author{Yang Song}\email{yangsong@pas.rochester.edu}
\author{Hanan Dery}\altaffiliation[Also at ]{Department of Electrical and Computer Engineering, University of Rochester, Rochester, New-York, 14627}
\affiliation{Department of Physics and Astronomy, University of Rochester, Rochester, New-York, 14627}
\begin{abstract}
We present a comprehensive theory of spin transport in a non-degenerate semiconductor that is in contact with multiple ferromagnetic terminals. The spin dynamics in the semiconductor is studied during a perturbation of a general, non-collinear magnetization configuration and a method is shown to identify the various configurations from current signals. The conventional Landauer-B\"{u}ttiker description for spin transport across Schottky contacts is generalized by the use of a non-linearized I-V relation, and it is extended by taking into account non-coherent transport mechanisms. The theory is used to analyze a three terminal lateral structure where a significant difference in the spin accumulation profile is found when comparing the results of this model with the conventional model.
\end{abstract}
\maketitle
\section{Introduction}
Hybrid semiconductor/ferromagnet material systems play a key role in
spintronics research \cite{Zutic_RMP04}. The motivation to study
these systems is twofold. First, computing technologies rely on the ability to easily tune the
carrier density in semiconductors. Second, the advances in storage applications rely on the ability to inject or extract spin
polarized electrons across interfaces between non-magnetic and
ferromagnetic materials \cite{Johnson_Silsbee_PRB87}-\cite{Prinz_Science98}. In the last decade, spin injection from ferromagnetic materials into
semiconductors has been showing a significant progress \cite{Zhu_PRL01}-\cite{Appelbaum_PRL07} together with a better understanding of the interface transport properties \cite{Butler_JAP97}-\cite{Honda_PRB08}. This progress has been accompanied with theoretical analysis of basic spin transport phenomena starting with the conductivity mismatch between a magnetic metal and a semiconductor
\cite{Johnson_Silsbee_PRB87,Schmidt_PRB00,Rashba_PRB00,Fert_PRB01,Smith_PRB01}, and continuing with effects
of electrical fields \cite{Yu_Flatte_long_PRB02, Albrecht_PRB03}, of lateral transport \cite{Dery_lateral_PRB06}, and of time dependent response \cite{Rashba APL02,Cywinski_APL06,Dery_Nature07}. In this paper we provide a comprehensive theory of time-dependent spin
transport in semiconductor/ferromagnet (SC/FM) systems. It studies
the potential and spin accumulation profiles in a general
non-collinear setup of magnetization directions. Two new aspects are provided in deriving the transport equations.  First, we elaborate on the quasi-neutrality approximation that simplifies the description of the drift diffusion equations. We show that the reasoning for this often-invoked approximation is different than what has been assumed since the late 1940's \cite{Shockley_BSTJ49}-\cite{Roosbroeck_PR61}. Second, we take into consideration the localization of electrons due to the doping inhomogeneity of typical SC/FM junctions
\cite{Hanbicki_APL02,Jonker_IEEE03,Adelmann_JVST05}. This leads to a change
of the canonical boundary conditions that rely on the
Landauer-B\"{u}ttiker formalism \cite{Schmidt_PRB00}-\cite{Dery_Nature07}. We use this model
to analyze lateral geometries which capture the vast majority of
spin injection experiments \cite{Johnson_Science93}-\cite{Crooker_PRB09}. The analysis considers the intrinsic capacitance of the SC/FM contacts and the
two-dimensional profiles of the electrical field and spin
accumulation \cite{Cywinski_APL06, Dery_Nature07}.

This paper is organized as follows. Sec.~\ref{sec:general formalism} presents a time dependent analysis
of spin transport in a bulk semiconductor region and across a SC/FM
junction. Sec.~\ref{sec:real_systems} discusses the modifications
that should be introduced in realistic systems. It deals with the
revision of the boundary conditions in forward biased junctions and
with their voltage bias limitations. In Sec.~\ref{sec:RD} we apply
our model to a non-collinear, three terminal planar geometry and we
quantify the time dependent readout across a capacitive barrier.
Sec.~\ref{sec:so} provides a summary. Descriptions of technical numerical procedures are provided in separate appendices.


\section{General Formalism}\label{sec:general formalism}
Theoretical analysis of spin injection from metals into
semiconductors shows that non-ohmic junctions are
necessary for the current to be polarized
\cite{Johnson_Silsbee_PRB87,Schmidt_PRB00,Rashba_PRB00,Fert_PRB01,Smith_PRB01}. More precisely, since
the spin-depth conductance of the semiconductor (conductivity
divided by spin-diffusion length) is much smaller than its metal
counterpart, for spin injection to occur the junction conductance
has to be similar or smaller than the semiconductor spin-depth
conductance. This spin injection constraint is easily achieved by an insulator
barrier or by the naturally formed Schottky barrier at the interface
between a metal and a semiconductor \cite{Tunneling_Phenomena,Sze}.
In the case of a thin tunneling barrier, the electrochemical
potential is discontinuous at the junction and as a result, the spin polarization of the
current is driven solely by the spin
selective transmission across the junction. The much larger conductivity and spin-depth
conductance in the ferromagnet render the spatial and spin dependence of its potential level negligible. Thus, in the following we describe the spin transport only in
the bulk semiconductor region and across the SC/FM junction whereas the ferromagnet is considered as a reservoir with a uniform potential level. We investigate in detail lateral systems that consists of ferromagnetic contacts on top of a non-degenerate semiconductor channel. When applicable, we rely on previous theoretical investigations of spin-transport in metals. These include both time dependent \cite{Zhang_PRB02}-\cite{Zhu_PRB08} and non-collinear \cite{Slonczewski_PRB89}-\cite{Xu_Nanotechnology08} aspects. In our analysis, we do not consider the effects of ballistic transport \cite{Brataas_PRL00}, of external magnetic fields \cite{Hernando_PRB00}, or of anisotropic spin relaxation \cite{Saikin_JPCM04}.

\subsection{Bulk Semiconductor} \label{sec:Bulk SC}

Macroscopic transport equations describe particle conservation and
current processes. These equations can be derived from the zeroth
and first moments of the dynamical Boltzmann transport equation
\cite{Valet_PRB93} and they provide a spatial and temporal
connection between spin dependent electron and current densities
($n_{\pm}(\mathbf{r},t) $ \& $j_{\pm}(\mathbf{r},t)$). The
accumulated (depleted) spin population at $(\mathbf{r},t) $ is directed in the $+$ ($-$) direction.
Using the relaxation time approximation, these derived transport equations are given by,
\begin{eqnarray}
\frac{\partial n_{s}}{\partial t} &=&
\frac{1}{q}\nabla \cdot \mathbf{J}_{s}\,
-\, \frac{n_{s}}{\tau_{s,s'}}\,
+\frac{n_{s'}}{\tau_{s',s}}\,, \label{eq:zeroth_moment} \\
\tau_{s,m}\frac{\partial
\mathbf{J}_{s}}{\partial t} &=&
qD_{s}\nabla n_{s}+\sigma_{s}\textbf{E}\,
-\mathbf{J}_{s}\,,\label{eq:first moment}
\end{eqnarray}
where the indices $s$ \& $s'$ denote either $(+,-)$ or $(-,+)$.
$q>0$ is the elementary charge and $\mathbf{E}$ is the macroscopic
electric field. The spin-dependent macroscopic parameters are the
spin-flip time from spin $s$ to spin $s'$ and vice versa ($\tau_{s,s'}$
\& $\tau_{s',s}$), the diffusion coefficients ($D_{s}$), the
conductivities ($\sigma_{s}$) and the momentum relaxation times
($\tau_{s,m}$). The current terms are eliminated by substituting the
continuity equation into the divergence of the current equation,
\begin{eqnarray}
 &\!& \!\!\!\!\!\!\!\frac{\partial n_{s}}{\partial t} + \frac{n_{s}}{\tau_{s,s'}} - \frac{n_{s'}}{\tau_{s',s}} = \mathbf{\nabla} \cdot \left(D_{s} \mathbf{\nabla} n_{s} \right) +\frac{1}{q}\mathbf{E} \cdot \nabla \sigma_{s}  \label{eq:n2A} \\
&\!&\!\!\!\!\!+ \! \left \{\!  \frac{\sigma_{s}}{q} \mathbf{\nabla}\! \cdot\! \mathbf{E} - \tau_{s,m} \!\left(  \frac{\partial^2 n_{s}}{\partial t^2}  + \frac{1}{\tau_{s,s'}} \frac{\partial n_{s}}{\partial t} -\!  \frac{1}{\tau_{s',s}} \frac{\partial n_{s'}}{\partial t}  \right) \right \} . \nonumber
\end{eqnarray}
At this phase, one can derive the dynamical spin dependent
drift-diffusion equation by applying a series of controlled
approximations after which the first line is rewritten in a more
compact form and the second line (curly brackets) is neglected
\cite{Schmidt_PRB00}-\cite{Dery_Nature07}. In non-degenerate and
homogeneous semiconductors, the diffusion constant and momentum
relaxation time are spin and position independent:
$D_{+}$=$D_{-}$=$D$ \& $\tau_{+,m}$=$\tau_{-,m}$=$\tau_m$. In
addition, the spin-flip times are equal and much greater than the
momentum relaxation time,
$\tau_{+,-}$=$\tau_{-,+}$=$2\tau_{sf}$$\,$$\gg$$\,$$\tau_m$.  We can
therefore accurately approximate the above equation as,
\begin{eqnarray}
\frac{\partial n_{s}}{\partial t} \! + \! \frac{n_{s}\!-\!n_{s'}}{2\tau_{sf}} \!= \!D \nabla^2 \!n_{s} \! +\! \nu \mathbf{E}\! \cdot \! \!\nabla n_{s}  \!
+ \! \frac{\sigma_{s}\!\mathbf{\nabla} \!\! \cdot \!\! \mathbf{E} }{q} \! - \! \tau_{m} \frac{\partial^2 n_{s}}{\partial t^2}    ,  \label{eq:n2B}
\end{eqnarray}
where $\nu$ denotes the mobility ($\sigma_{s} \equiv q \nu n_{s}$).
Zhu \textit{et al.} have studied the wave-like behavior due to the
second order time derivative in magnetic metallic systems (vanishing
$\mathbf{E}$ terms) \cite{Zhu_PRB08}. They have shown that this
effect becomes significant at time scales shorter than $\tau_m$. On
the other hand, if the interest is in semiconductors and in much
longer time scales then a different approach is needed. First, we
define the spin polarization along the $\pm$ axis and the charge
accumulation,
\begin{eqnarray}
p = \frac{n_{+}-n_{-}}{n_0},  \,\,\,\,\,\, \rho = \frac{n_{+}+n_{-} - n_0}{n_0}.  \label{eq:PsPc}
\end{eqnarray}
$n_0$ denotes the electron density in the conduction band due to the background doping. Taking into account the Poisson equation as well as the difference and sum of Eq.~(\ref{eq:n2B}) with its corresponding equation ($+ \leftrightarrow -$) one gets,
\begin{eqnarray}
\!\!\!\!\! \frac{\partial p}{\partial t} \!+\! \tau_{m} \frac{\partial^2 p}{\partial t^2} \!\! & = & \!\!- p \left(\frac{1}{\tau_{sf}}\! + \! \frac{\rho}{\tau_{d}} \right) + \! D \nabla^2  p\! + \nu \mathbf{E}\! \cdot \!\mathbf{\nabla}p , \label{eq:diffPs}\\
\frac{1}{\tau_{m}}\frac{\partial \rho}{\partial t} \!+\!  \frac{\partial^2 \rho}{\partial t^2} \!\! & = & \!\!- \omega_p^2 \rho ( 1\!+\!\rho )  +   \frac{D}{\tau_m} \nabla^2 \rho \!+ \frac{q}{m_{sc}} \mathbf{E} \!\cdot \! \mathbf{\nabla}\rho ,  \label{eq:diffPc} \\
\mathbf{\nabla} \cdot \mathbf{E}   & = & -\frac{\rho}{\nu \tau_d} \,\,.  \label{eq:Poisson}
\end{eqnarray}
$\tau_d$ is the dielectric relaxation time defined by the ratio between
the static dielectric constant and the total conductivity,
$\epsilon_{sc}/\sigma_0$, where $\sigma_0=q \nu n_0$. The plasma frequency is defined by $\omega_{p} \equiv 1/\sqrt{\tau_d\tau_m}=\sqrt{q^2 n_0/m_{sc}\epsilon_{sc}}$ where $m_{sc}$ is the effective mass of the electron in the semiconductor. Since in most cases the interest is in time scales much longer than the momentum relaxation time,  it is common to neglect the wave-like behavior already when describing the current components ($\tau_{s,m}$=0 in Eq.~(\ref{eq:first moment})). The resulting charge dynamics is then described by a diffusion equation which in the linear regime of small charge perturbations reads $\partial \rho/\partial t  = - \rho/\tau_d  +  D \nabla^2 \rho$. The argument for invoking the quasi-neutrality approximation is then that any local charge imbalance ($\rho
\neq 0$) is being screened out within a time scale of the order of
$\tau_d$. This is a widely used argument, whose origin can be traced back to the seminal works on bipolar transport in homogeneous semiconductors \cite{Shockley_BSTJ49}-\cite{Roosbroeck_PR61}. However, by keeping the wave-like terms then the decay of $\rho$ is actually governed by $\tau_m$ and it is nearly independent of $\tau_d$. This is a manifestation of the finite propagation velocity which also results in an oscillatory behavior during the relaxation. It is still justifies, however, to
assign $\rho$=0 in Eqs.~(\ref{eq:diffPs}) and (\ref{eq:Poisson}) but the argument should refer to interest in time scales much longer than $\tau_{m}$. This statement is general and should not change qualitatively if the original transport equations ((\ref{eq:zeroth_moment}) \& (\ref{eq:first moment})) are derived without employing the relaxation-time approximation. Moreover, the
exclusion of spin dependent parameters in Eqs.~(\ref{eq:diffPc}) and
(\ref{eq:Poisson}) also suggests that the charge dynamics is
general. For example, similar charge dynamics describes bipolar
transport in homogeneous semiconductors. The charge accumulation is then due to
deviation of hole and electron densities from their local
equilibrium, $\rho=(\Delta n -\Delta p)/(n_0+p_0)$. In a different
publication we will revisit the widely used quasi-neutrality
concept, and elucidate the true nature of ultra-fast charge dynamics
in various systems \cite{Yang_neutrality}. Here, we provide
an example that investigates the applicability of the adiabatic
approximations ($\tau_m=0,\rho=0$) in deriving the spin dependent
transport equations in semiconductors. One should recall, however, that this classical approach (Eqs.~(\ref{eq:diffPc})~\&~(\ref{eq:Poisson})) neglects the effect of dynamical screening. At relatively low electron densities (e.g., non-degenerate semiconductors) this classical description is accurate since the screening length is larger or comparable to the mean free path.

We study the charge accumulation evolution $\rho(\mathbf{r},t)$
after a disturbance that locally breaks the charge neutrality in an
overall neutral bulk semiconductor. The charge current density at
the boundaries of the system is $J_0$ at all times. The dynamical response is similar for all systems in which the external electric field is negligible compared with the initial built-in electric field due to the charge imbalance ($J_0$ is smaller than some critical current density). For simplicity the analysis proceeds with $J_0=0$. We assume the disturbance happens far from the system
boundaries and that it is spherically symmetric. Based on these
conditions, the charge evolution and the electric field
possess a spherical symmetry. We choose a representative initial
charge profile due to the disturbance,
$\rho(r,t=0)=\rho_0\left[\exp(-\frac{r^2}{2 \Delta^2})-C \left(
\exp(-\frac{(r-2 \Delta)^2}{2 \Delta^2})+\exp(-\frac{(r+2
\Delta)^2}{2 \Delta^2}) \right)\right]$. $C$ is a constant chosen to keep the integrated space charge zero and $\rho_0$ reflects the initial intensity of the disturbance at the
center. The spatial extent of the disturbance is determined by $\Delta$. For the second needed initial information we set $\partial\rho(r,t)/\partial{t}|_{t=0}=0$ (this choice does not qualitatively change the following discussion).

\begin{figure}
\includegraphics[width=8cm]{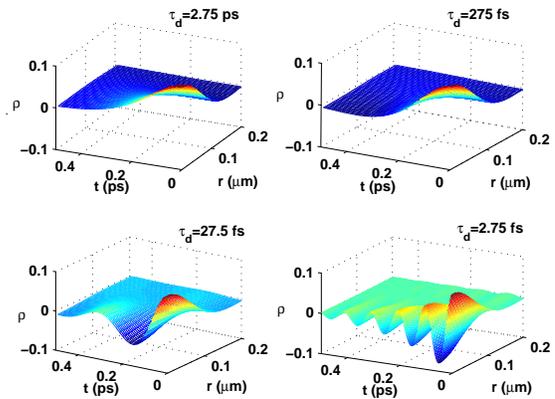}
\caption[quasi_neutrality] {(Color online) Charge disturbance evolution and propagation
$\rho(r,t)$. (a)-(d) are the results for different dielectric relaxation times, $\tau_d=\{$2.75~ps,~275~fs,~27.5~fs~\&~2.75~fs$\}$ which corresponds, respectively, to equilibrium electron densities of $n_0=\{$10$^{15}$,~10$^{16}$,~10$^{17}$~\&~10$^{18}\}$~cm$^{-3}$. The relaxation dynamics is similar and only slightly affected by the dielectric relaxation time. The decay time is about $2\tau_m$=200~fs. At high densities, the oscillations are at the plasma frequency.} \label{fig:quasi_neutrality}
\end{figure}

Fig.~\ref{fig:quasi_neutrality} shows the evolution of a charge disturbance whose peak intensity is $\rho_0=0.1$ and its spatial extent is $\Delta$=60~nm. We consider a room temperature, non-degenerate n-type GaAs with a momentum relaxation time of $\tau_m=100$~fs. The resulting mobility and diffusion constant are, respectively, $\nu\,\approx\,$2600~cm$^2$/V$\cdot$s and $D\approx$68~cm$^2$/s. Panels (a)-(d) show, respectively, the evolution with these parameters for doping densities of $n_0 \sim 10^{15},\,10^{16},\,10^{17},\,\&\,10^{18}$~cm$^{-3}$. The resulted dielectric relaxation time is changed over three orders of magnitude ($\tau_d=\epsilon_{sc}/(q \nu n_0)$). In spite of the large changes in $\tau_d$, the decay time is about 2$\tau_m$=200~fs in all cases. The exp$(-t/2\tau_m)$ decay is a universal behavior if $2\tau_m\omega_p > 1$ \cite{Yang_neutrality}. The results also show a clear oscillatory behavior at shorter dielectric relaxation times where the oscillation frequency matches the plasma frequency, $\omega_{p}$. To understand this behavior we consider the Fourier transform of the initial disturbance. Its effective width is $\sim 1/\Delta$ and its coherence time scale is defined by $\tau_c \equiv \Delta^2/D$. If the initial disturbance is relatively wide such that, $\tau_c>> \tau_d$, then the oscillations are governed by the (central) plasma frequency. The oscillatory behavior is damped when $\tau_c \leq \tau_d$ due to the destructive interference between the wavevector components of the disturbance. Studying these and other effects (e.g., the role of momentum relaxation time, the non-linear terms, confined and open systems) are beyond the scope of this paper and will be studied elsewhere \cite{Yang_neutrality}.

To summarize the quasi-neutrality aspect, if the interest is in spin phenomena at time scales much longer
than $\tau_m$ (and not $\tau_d$), then it is accurate to apply the adiabatic approximations
($\tau_m=0,\rho=0$), and get a divergence-free electric field and a
linear dynamical spin-drift-diffusion equation,
\begin{eqnarray}
\nabla\cdot \mathbf{E}&=&0\,\,, \label{eq:Edivergence0}\\
\frac{1}{D}\frac{\partial p}{\partial t} +  \frac{p}{\ell_{sf}^2} & = &\nabla^2  p  + \frac{1}{V_T} \mathbf{E}  \cdot \mathbf{\nabla}p \,\,, \label{eq:LinDiffPs1D}
\end{eqnarray}
where $\ell_{sf} = \sqrt{D \tau_{sf}}$ is the spin-diffusion length
and the Einstein relation was invoked, $D/\nu= V_T \equiv k_BT/q$.

The spin dependent electrochemical potential, $\mu_{\pm}$, is also
an important transport quantity which has the following relation
with $p$,
\begin{eqnarray}
\frac{\mu_{\pm}(\mathbf{r},t)}{k_BT}  = \frac{\mu_0}{k_BT} +  \ln\left( 1 \pm p(\mathbf{r},t) \right)\, ,  \label{eq:mu}
\end{eqnarray}
where $\mu_0=\mu_c -q\phi(\mathbf{r},t) $ denotes the spin
independent part defined by the sum of a constant chemical potential
and the electrical potential. The latter is driven by the applied
bias voltage and is related to the electrical field via $\textbf{E}=
-\mathbf{\nabla} \phi$. The logarithmic term refers to the
non-degenerate case (with $\rho=0$).

To this point, we have treated the spin polarization in the channel,
$p$, as a scalar which implicitly relies on the assumption that the
net spin has a fixed direction throughout the semiconductor channel.
This description is valid in collinear systems at which the
magnetization directions in all of the ferromagnetic elements share a common (easy) axis. In a more general, non-collinear
configuration the boundary conditions impose a change in the direction
of the spin polarization during the transport in the channel. For a
general coordinate system in spin space, the spin dependent electron
density is described by a $2$x$2$ matrix,
\begin{eqnarray}
\hat{\mathfrak{n}} (\mathbf{r},t) = \frac{n_0}{2} \left( \hat{I} + \mathbf{p}(\mathbf{r},t)\cdot\hat{\boldsymbol{\sigma}} \right)\,\,, \label{eq:nMat}
\end{eqnarray}
where $\hat{\boldsymbol{\sigma}}$ is the Pauli matrix vector and
$\mathbf{p}$ has the magnitude $p$ along the $+$ direction as
defined in Eq.~(\ref{eq:PsPc}). Using this notation and repeating
the analysis, the components of the
spin-drift-diffusion equation read,
\begin{eqnarray}
\frac{1}{D}\frac{\partial p_i }{ \partial t } +  \frac{p_i}{\ell_{sf}^2} = \sum_{j} \left( \frac{\partial^2 p_i}{ \partial x_j^2}  + \frac{E_j}{V_T}  \frac{\partial p_i}{ \partial x_j} \right) \,\,, \label{eq:LinDiffPs}
\end{eqnarray}
where $i$ ($j$) enumerates the $x$, $y$ and $z$ coordinates in spin
(real) space. According to Eq.~(\ref{eq:first moment}), the
components of the charge current density (vector) and of the spin
current density (second-rank tensor) are,
\begin{eqnarray}
J_j & = & \sigma_0 E_j \,\,. \label{eq:J_0} \\
\mathfrak{J}_{i,j} & = & {\sigma_0} \left( V_T \frac{\partial p_i}{
\partial x_j}  + E_j  p_i \right) \,\,. \label{eq:Js}
\end{eqnarray}
These expressions are valid if the frequency of the applied
electrical signal is much smaller than $1/\tau_m$.

\subsection{SC/FM Junction} \label{sec:SC/FM}

The description of transport is complete when the spin polarized
currents across the SC/FM junctions are expressed in terms of the
spin polarization vector at the semiconductor side of the junction,
$\mathbf{p}(\mathbf{r}_j,t)$. We follow the notation by Brataas
\textit{et al.} \cite{Brataas_PRL00, Brataas_EPJB01}, and use
Eqs.~(\ref{eq:mu}) \& (\ref{eq:nMat}) to write the population
distribution matrices on both sides of the junction,
\begin{eqnarray}
\hat{f}_{sc}(\varepsilon) &=&  e^{(\mu_0-\varepsilon)/k_BT} \left( \hat{I} + \mathbf{p}\cdot\hat{\boldsymbol{\sigma}} \right)\,\,, \label{eq:f_sc_Mat} \\
\hat{f}_{fm} &=& \left( 1+ e^{(\varepsilon-\mu_0+qV)/k_BT} \right)^{-1} \hat{I}  \,\,, \label{eq:f_fm_Mat}
\end{eqnarray}
where $\varepsilon$ denotes the energy. As mentioned before, due to the conductivity mismatch the ferromagnetic side is a reservoir with a constant chemical potential, $\mu_0 -qV$, where $\mu_0$ (Eq.(\ref{eq:mu})) is evaluated at the semiconductor side of the junction and $V$ is the voltage drop across the SC/FM junction. $V>0$ ($V<0$) denotes forward (reverse) bias voltage in which electrons flow into (from) the ferromagnetic contact. For compact notation, the boundary conditions of each SC/FM junction are written in a spin coordinate system at which the $z$-axis is collinear with the majority spin direction of the corresponding ferromagnetic contact. With this simplification, the reflection matrices are diagonal,
\begin{eqnarray}
\hat{r}_{sc }(\varepsilon_{\perp},V) = \left(\!\begin{array}{cc}
                                         r_{\uparrow} & 0 \\
                                         0 &  r_{\downarrow}
                                       \end{array} \!\! \right) \,\,, \,\, \hat{r}_{fm }(\varepsilon_{\perp},V)  = \left( \! \begin{array}{cc}
                                         \tilde{r}_{\uparrow} & 0 \\
                                         0 &  \tilde{r}_{\downarrow}
                                       \end{array} \!\! \right). \label{eq:r_mat}
\end{eqnarray}
The reflection coefficients in the left (right) matrix are of
electrons from the semiconductor (ferromagnetic) side of the
junction. For a given material system, these coefficients vary with
the voltage drop and with the longitudinal energy,
$\varepsilon_{\perp}$, which denotes the impinging energy of
electrons due to their motion toward the SC/FM interface. The up and
down arrows denote, respectively, the majority and minority spin
directions in the ferromagnetic contact where we have set the $+z$
direction parallel to the majority direction. By using the
Landauer-B\"{u}ttiker formalism, the tunneling current density across the
SC/FM junction is given by \cite{Brataas_PRL00,Brataas_EPJB01,Ciuti_PRL02},
 \begin{eqnarray}
\hat{J}(V) &=& \int_{0}^{\infty} d \varepsilon \hat{j}(\varepsilon)
= \frac{q}{h} \int_{0}^{\infty} d \varepsilon
\int_{0}^{k_{\varepsilon}}  \frac{d^2 k_{\parallel}}{(2\pi)^2}
\label{eq:J_mat_general} \\ &\!& \left\{ \left[ \hat{f}_{fm} -
\hat{r}_{fm} \hat{f}_{fm} \hat{r}^{\dag}_{fm}  \right]   -    \left[
\hat{f}_{sc} - \hat{r}_{sc} \hat{f}_{sc} \hat{r}^{\dag}_{sc}
\right]  \right\}. \nonumber
\end{eqnarray}
The first (second) term in square brackets is related to the
transmitted current from the ferromagnet (semiconductor) due to
electrons whose total and longitudinal energies are $\varepsilon$
and $\varepsilon_{\perp}$, respectively. The zero energy refers to the bottom of the semiconductor conduction band. The inner integration is
carried over transverse wavevectors due to a motion in parallel to
the SC/FM interface and its upper integration limit,
$k_{\varepsilon}$, denotes the wavevector amplitude of an electron
with energy $\varepsilon$. In the chosen spin coordinate system, the
transmitted spin current from the ferromagnetic side is non-zero
only along the $z$ direction. Its spin-up and spin-down components
are proportional, respectively, to ($1-|r_{\uparrow}|^2$) and
($1-|r_{\downarrow}|^2$), where we have rendered the fact that
$|r_{\uparrow\,(\downarrow)}|^2=|\tilde{r}_{\uparrow\,(\downarrow)}|^2$.
The  transmitted current from the semiconductor side, on the other
hand, includes off-diagonal mixed terms that are proportional to
$r_{\uparrow}r^{\ast}_{\downarrow}$. This is the case when
$\mathbf{p} \nparallel \mathbf{z}$ due to the flow of electrons
from/into ferromagnetic contacts which has non-collinear
magnetization directions and that are located within about a spin-diffusion length. By substituting
Eqs.~(\ref{eq:f_sc_Mat})-(\ref{eq:r_mat}) into
Eq.~(\ref{eq:J_mat_general}) we write the energy resolved tunneling
current density matrix,
\begin{eqnarray}
q\hat{j}(\varepsilon) & = &  \left( 1+ e^{(\varepsilon-\mu_0+qV)/k_BT} \right)^{-1}\left( \begin{array}{cc}
                                         g_{\uparrow} & 0  \\
                                         0 &  g_{\downarrow}
                                       \end{array} \right)  \label{eq:interface current per energy} \\ &  - & e^{(\mu_0-\varepsilon)/k_BT} \left( \begin{array}{cc}
                                         g_{\uparrow}(1+p_z) & g_{\uparrow \downarrow}(p_x - ip_y) \\
                                         g_{\uparrow \downarrow }^{*} (p_x + ip_y) &  g_{\downarrow}(1-p_z)
                                       \end{array} \!\!\right). \nonumber
\end{eqnarray}
The $p_i$ components of the spin polarization vector are evaluated
at the semiconductor side of the junction, and the direct and mixing
conductances (per unit area) are given by,
\begin{eqnarray}
\!\!\!\!\!\!\!\!\!\!\!\!\!\!\!g_{\uparrow(\downarrow)}\!(\varepsilon, V) &=& \frac{q^2}{h} \int_0^{k_{\varepsilon}} \frac{d^{2} k_{\parallel } }{(2\pi )^{2} } \left(1- |r_{\uparrow(\downarrow)} |^2 \right) \,, \label{eq:conductance per energy}\\
\!\!\!\!\!\!\!\!\!\!\!\!\!\!\!g_{\uparrow \downarrow }\!(\varepsilon, V) &=& \frac{q^{2} }{h} \int_0^{k_{\varepsilon}} \frac{d^{2} k_{\parallel } }{(2\pi )^{2} } \left(1-r_{\uparrow } r_{\downarrow } ^{*} \right) \,. \label{eq:mixing conductance per energy}
\end{eqnarray}
In this writing, the $\varepsilon_{\perp}$ dependence of the
reflection coefficients is resolved from $\varepsilon$ \&
$k_{\parallel }$. The analysis is further simplified if we assume
that the potential level in the ferromagnetic contact lies beneath
the edge of the semiconductor conduction band, $\mu_0-qV < 0$.
Since $\mu_0<0$ in a non-degenerate semiconductor , this
assumption holds for any forward bias conditions and for relatively
low reverse bias conditions. Thus, we can consider only the
Boltzmann tail of the ferromagnetic Fermi-Dirac distribution (first term in
Eq.~(\ref{eq:interface current per energy})). In this
regime, the components of the tunneling current density across the SC/FM
boundaries are compactly described by,
\begin{align}
J_{\alpha}(V)&=V_T G\left(e^{-V/V_T} - 1 - F p_z \right)\,,\label{eq: interface j0}\\
\mathfrak{J}_{z,\alpha}(V)&=V_T G\left( F\left(e^{-V/V_T} - 1\right) - p_z  \right)\,,\label{eq: interface jsz}\\
\mathfrak{J}_{y,\alpha}(V)&=2V_T\left( \text{Im}[G_{\uparrow \downarrow }]  p_x\!\!-\! \text{Re}[G_{\uparrow \downarrow }  ]p_y\right)\,,\label{eq: interface jsy}\\
\mathfrak{J}_{x,\alpha}(V)&=2V_T\left(- \text{Re}[G_{\uparrow \downarrow }]p_x\!\!-\! \text{Im}[G_{\uparrow \downarrow }  ]p_y\right)\,.\label{eq: interface jsx}
\end{align}
The subscript $\alpha$ is a real space coordinate directed along the normal of the SC/FM interface. The total and mixing macroscopic conductances ($G$ \& $G_{\uparrow \downarrow }$) and the finesse ($F$) of the junction are given by,
\begin{eqnarray}
G &=& \frac{1}{qV_T} \int_{0}^{\infty} \!\! d \varepsilon e^{(\mu_0-\varepsilon)/k_BT}\left( g_{\uparrow}\!(\varepsilon,V)+g_{\downarrow}\!(\varepsilon,V) \right),  \nonumber \\
G_{\uparrow \downarrow } & = & \frac{1}{qV_T} \int_{0}^{\infty} \!\! d \varepsilon e^{(\mu_0-\varepsilon)/k_BT} g_{\uparrow \downarrow}\!(\varepsilon,V) , \label{eq:Gmix} \\
F &=& \frac{1}{qV_T G} \int_{0}^{\infty} \!\! d \varepsilon e^{(\mu_0-\varepsilon)/k_BT}\left( g_{\uparrow}\!(\varepsilon,V)-g_{\downarrow}\!(\varepsilon,V) \right) . \nonumber
\end{eqnarray}

The reflection coefficients that appear in the boundary currents can
be either extracted from carefully designed experiments
\cite{Kawakami_Science01,Epstein_PRB02,Li_PRL08}, or calculated by various
techniques. In this paper, we use an effective mass single band
model \cite{Ciuti_PRL02}, \cite{Osipov_PRB04}-\cite{Smith_PRB08}. Details of the calculation are given in Appendix~\ref{sec:interface current}. We mention that this model does not include the predicted effects of Ab-initio
calculations \cite{Butler_JAP97}-\cite{Honda_PRB08}.  These effects
may become crucial in ideal SC/FM interfaces and concern
spin-filtering mechanisms or the presence of interfacial bands.

\section{Realistic Modeling} \label{sec:real_systems}

There are several restrictions and mechanisms that should be
considered in a realistic modeling of spin transport in biased SC/FM
hybrid systems. The use of Poisson and linear spin-drift-diffusion
equations ((\ref{eq:Edivergence0}) \& (\ref{eq:LinDiffPs})) inside
the homogenous bulk semiconductor is a reliable macroscopic
description as long as ballistic effects are not important. However,
three crucial aspects about the boundary currents need to be
addressed. The first relates to the inhomogeneous doping profile at
the Schottky barrier and it results in a change of the boundary
conditions. The second aspect relates to the bias voltage across
SC/FM junctions and it limits the applicability of the boundary
conditions to a finite bias voltage range. The third relates to the
intrinsic capacitance of the Schottky barriers and it plays a role
in the dynamics.

\subsection{Inhomogeneous doping profile} \label{sec:doping}

At low bias voltages, the width of the barrier should be about or
less than 10~nm in order to suppress thermionic currents at room
temperature (see Appendix~\ref{sec:interface current}). Therefore, the doping
concentration at the Schottky junction should be highly degenerate,
$n_{sb} \gg $10$^{18}$~cm$^{-3}$. If the bulk of the semiconductor
is non-degenerate ($n_0 \ll $10$^{18}$~cm$^{-3}$) then the result is
a strongly inhomogeneous doping profile between these regions
\cite{Hanbicki_APL02,Jonker_IEEE03,Adelmann_JVST05}. The need for a much
lower carrier density in the bulk region, stems from the condition
of optimal spin accumulation density: the barrier conductance is of
the same order of magnitude as the semiconductor spin-depth
conductance $G \sim \sigma_0/l_{sf}$
\cite{Fert_PRB01,Dery_lateral_PRB06}. Since $G$ is relatively small
even with a $\sim$10~nm wide barrier, the optimal `spin-impedance'
condition can be met with a bulk region that is moderately doped
(low $\sigma_0$) for which the spin-diffusion length, $l_{sf}$, is
relatively large. Thus, for the spin accumulation density to be
non-negligible the background doping densities are typically such
that $n_{0}\,< \,10^{17}$~cm$^{-3}$. As a result of the doping
inhomogeneity between the bulk and the barrier regions, a potential
well is likely to be created between these regions
\cite{Zachau_SSC86}-\cite{Shashkin_Semiconductors02}. Even with a
careful doping design at which there is no well in equilibrium, at
forward bias when less electrons need to be depleted from the
semiconductor, the well creation is inevitable. The spin related
effects in this potential well may contribute to the spin
accumulation in the bulk semiconductor region \cite{Dery_PRL07,
Li_APL09}.

\begin{figure}
\includegraphics[width=8cm]{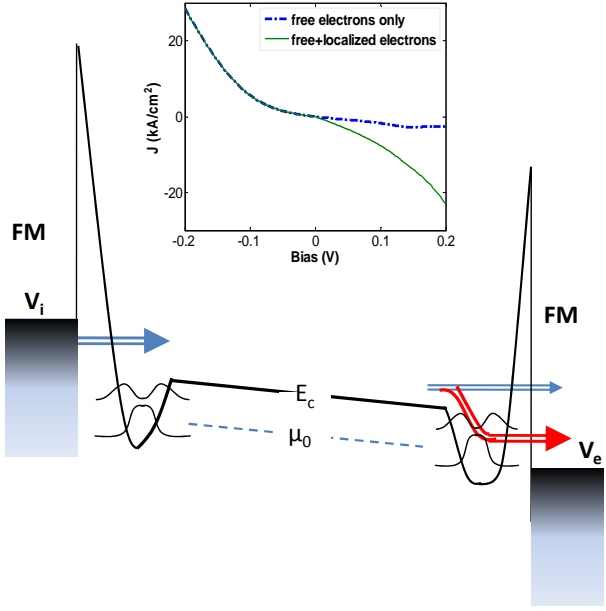}
\caption[discretized_SC] {(Color online) A scheme of the conduction band profile of
a one-dimensional FM/SC/FM structure. In forward bias, the potential
well contributes to the current across the interface. Inset: room-temperature J-V
curve across an n-type GaAs/Fe junction where the semiconductor bulk
doping is 10$^{16}$~cm$^{-3}$ and the interface doping is
$2\times10^{19}$~cm$^{-3}$.                  }
\label{fig:well_structure}
\end{figure}

Fig.~\ref{fig:well_structure} shows the conduction band profile
across a one-dimensional biased system (calculation details in
Appendix \ref{sec:interface current}). The left and right barriers
denote, respectively, the reverse (spin injection) and forward (spin
extraction) biased junctions. Across the reverse biased junction, most of the injected hot
electrons overshoot this region (although their reflection
coefficients are slight modified by the potential well). On the other hand, in the forward
direction, we should consider a new transport mechanism which involves the escape
of spin polarized electrons from the potential well into the
ferromagnetic contact. This process accompanies the previous
mentioned process of free electrons tunneling
(Eq.~(\ref{eq:J_mat_general})). Therefore, free electrons from the
bulk semiconductor region can either tunnel directly into the
ferromagnetic metal or feed the potential well when its localized
electrons escape. To quantify the spin dependent currents that flow
to the potential well, we consider three mechanisms. The first is
the capture time of a free electron into the potential well. This is
a spin independent time scale of the order of hundreds of fs and it
is governed by spin conserving phonon-carrier or carrier-carrier
scattering processes \cite{Deveaud_APL88}-\cite{Dery_PRB03}. The
second mechanism is the spin relaxation in the potential well and
its time scale is of the order of a few ps in III-V semiconductor
quantum wells \cite{Malinowski_PRB00}. The third mechanism is the
spin dependent escape time of electrons from the \textit{i}$^{th}$
localized state of the potential well into the ferromagnetic
contact. Its order of magnitude can be calculated by a WKB method,
\begin{eqnarray}
\frac{1}{\tau_{esc,i}} = \frac{1}{\tau_{esc,i, \uparrow}} + \frac{1}{\tau_{esc,i, \downarrow}}  & \approx &
\frac{\omega}{2\sqrt2} \cdot e^{-2\frac{\phi_B +\mu_0 - qV - E_i}{\hbar\omega}}, \,\,\,\,\,\,\, \label{eq:well_1}\\
\omega^2 &=& \frac{q^2 n_{sb}}{m_{sc} \epsilon_{sc}}.\label{eq:well_2}
\end{eqnarray}
$\phi_b$ is the difference between the work function of the metal and the affinity of the semiconductor (if needed it can also factor the pinning of the Fermi level). The barrier height from the conduction band of the semiconductor is denoted by $\phi_B - \mu_0- qV$. The localization energy is denoted by $E_i$ (see Appendix~\ref{sec:interface current} for its calculation) and $\omega$ corresponds to the parabolic curvature of the conduction band at the barrier
region. The escape time scale is highly sensitive to the doping level of
the Schottky region. For example, it increases from 28~ps to 11~ns
when the doping is reduced from $n_{sb} \sim 2\times 10^{19}$ to
$n_{sb} \sim 7\times 10^{18}$. These values are calculated by using
GaAs bulk parameters, $m_{sc}=0.067 m_0$ and
$\epsilon_{sc}=1.16\times 10^{-12}$~F/cm and a typical value of $\phi_B +\mu_0 - qV - E_i = 0.7$~eV. This process is spin dependent
since the escape rates are proportional to the inverse of the
wavevector in the ferromagnetic side \cite{Dery_PRL07},
\begin{eqnarray}
\frac{\tau_{esc, \downarrow}}{\tau_{esc, \uparrow}} \approx \frac{k_{fm,\downarrow}}{k_{fm,\uparrow}}.
\end{eqnarray}
The faster escape rate of electrons with smaller wavevector (e.g.,
minority electrons in iron) provides a way to distinguish this
effect from the delocalized electron tunneling whose spin
polarization is opposite \cite{Crooker_Science05}. Due to the large
differences between these time scales, $\tau_{cap} \ll \tau_{s,well}
\ll \tau_{esc}$) we can assume that (I) every electron that escapes
from the potential well into the ferromagnet is being replenished
immediately by an electron with the same spin from the bulk region
and (II) inside the potential well the spin polarization is
negligible, $\mathbf{p}^{2D} \approx 0$, and as a result only the
total current density, $J^{2D}$,  and the spin current density in the $z$-spin
coordinate, $\mathfrak{J}^{2D}_{z,\alpha}$ are non-zero. According
to our coordinate system, the +$z$ coordinate denotes the majority
spin direction in the ferromagnet. In analogy with the boundary
conditions of delocalized electrons in Eqs.~(\ref{eq: interface
j0})-(\ref{eq: interface jsx}), we see that even when $p_z$=0 the
spin current density in this direction is non-zero if $F \neq$0. The role of
the finesse in the localized case is played by the spin-dependent
escape times. To comply with this physical picture, we add
phenomenological terms to the boundary conditions of the forward
biased junction,
\begin{eqnarray}
J_{\alpha}(V)&=&V_T G\left(e^{-V/V_T} - 1 - F p_z \right)    +  J^{2D} \,,  \label{eq: interface j0j_total}\\
\mathfrak{J}_{z,\alpha}(V)&=& V_T G\left( F\left(e^{-V/V_T} - 1\right) - p_z  \right)  + \mathfrak{J}^{2D}_{z,\alpha} \,.\,\,\,\,\,\,\, \label{eq: interface jsz_total}
\end{eqnarray}
The contributed current density from the potential well is given by,
\begin{eqnarray}
 J^{2D} &=& -\frac{1}{2} q \sum_i \frac{\tilde{n}_i}{\tau_{esc,i}}\,,  \label{eq:j0_2D}\\
\mathfrak{J}^{2D}_{z,\alpha} &=& F^{2D}J^{2D} \,,  \label{eq:jz_2D}\\
F^{2D} &=&  \frac{\tau_{esc, \downarrow}-\tau_{esc, \uparrow}}{\tau_{esc, \uparrow}+\tau_{esc, \downarrow}}\,. \label{eq:F2D}
\end{eqnarray}
$\tilde{n}_i$ denotes the (bias dependent) two-dimensional density
of electrons that are not Pauli blocked in the \textit{i}$^{th}$
localized state. The energies of these electrons are above the
potential level in the ferromagnetic contact and thus they can
contribute to the escape process.  The $1/2$ factor denotes the fact
that the spin polarization in the potential well is zero
($n_{i,+z}=n_{i,-z}=\tilde{n}_i/2$). The $x$ and $y$ components of
the localized spin current densities are zero ($p_x^{2D}=p_y^{2D}=0$) and
thus only the free electrons contribute in these directions
(Eqs.~(\ref{eq: interface jsy}) \& (\ref{eq: interface jsx})).

Incorporating the potential well contribution to the current density is the
only way that one can fit experimental $J-V$ curves. Using the
Landauer-B\"{u}ttiker formalism to calculate the total tunneling
current density of free electrons (the trace of
Eq.~(\ref{eq:J_mat_general})) shows that at low bias conditions the
current density is exponentially larger in the reverse direction ($J(-|V|)
\gg J(|V|)$ if $|V|$$\leq$0.2). This is shown at the inset of
Fig.~\ref{fig:well_structure} for $n_{sb} =
2\times$10$^{19}$~cm$^{-3}$ and $n_{0} = $10$^{16}$~cm$^{-3}$ which
are the typical experimental values in Fe/GaAs material systems
\cite{Hanbicki_APL02}-\cite{Adelmann_JVST05},\cite{Crooker_Science05}-\cite{Lou_NaturePhys07}. However, the total experimentally measured $J-V$
curves show that the forward bias supports larger current densities than the
reverse bias throughout this voltage range. Without fitting
parameters, this contradiction is settled by including the potential
well contribution \cite{Dery_PRL07}. There is an important
consequence from this analysis. The escape current of localized
electrons and the tunneling current of free electrons can result in
opposite spin polarizations. Therefore, one can engineer their
relative fraction by the doping profile \cite{Li_APL09}. For the
highly doped interface, the escape current dominates the
forward $J-V$ curve and as a result the spin polarization in the
semiconductor is always along the majority spin direction (assuming that
the semiconductor bulk region has 0 net spin before the
injection or extraction). We will revisit this point in
Sec.~\ref{sec:RD}.

In this paper, we assume that in reverse bias conditions the injected electrons overshoot the
potential well. This approximation is accurate at very low temperatures where localized electrons cannot gain enough energy in order to pop-out of the well. However, at room temperature there is another transport mechanism to consider. Electrons from the ferromagnetic contact can tunnel into unpopulated localized states in the potential well and then to pop-out into the bulk region by absorption of a phonon or by electron-electron scattering. It should remain clear, however, that throughout the reverse bias range, the injection into free bulk states is the dominant mechanism due to the lower barrier that is involved in this process. Thus, in the following simulations we consider the localized electrons only in forward bias where the escape current can become the dominant transport mechanism.

\subsection{Bias voltage limitations} \label{sec:Bias voltage limitations}

Complementary magneto-optical Kerr spectroscopy and electrical Hanle
measurements show that the spin polarization is appreciable only over a moderate bias voltage range both in the forward and reverse directions \cite{Crooker_PRB09}. In forward bias, there are two reasons that limit the spin polarization of extracted electrons with increasing the voltage across the SC/FM junction.
First, at large positive voltages electrons can tunnel into
ferromagnetic states above the Fermi energy where new bands with
smaller or zero spin polarization exist \cite{Chantis_PRL07}, \cite{Osipov_PRB04}. This effect can
be incorporated by calculating the reflection coefficients with
additional bands above the Fermi level of the ferromagnet. Second, the barrier height is lowered with forward bias and thus the conductance increases exponentially. The spin accumulation in
the semiconductor channel disappears when the barrier conductance significantly exceeds the spin-depth conductance of the
semiconductor ($G \gg \sigma_0/\ell_{sf}$). At this bias regime, the
voltage drop across the junction is negligible and as a result the
electrochemical potential splitting between $\mu_+$ and $\mu_-$
becomes negligible. Both mentioned restrictions limit the
ability to achieve spin selective extraction of free and
localized electrons (first and second terms of Eq.~(\ref{eq: interface jsz_total}), respectively).

\begin{figure}
\includegraphics[width=5cm]{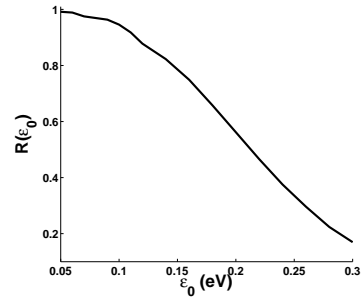}
\caption{The ratio between the net spin after energy thermalization and the net injected spin as a
function of the initial injected energy in a 10$^{16}$~cm$^{-3}$ n-type
GaAs. The method of calculation is presented in Ref.~\cite{Song_CondMat09}.}
\label{fig:hot_electrons}
\end{figure}

In reverse bias conditions, increasing the voltage amplitude limits
the spin polarization of injected electrons due to a transport
across a wider depletion region with enhanced electric field
\cite{Albrecht_PRB03, Saikin_JPCM04, Weng_PRB04} and due to an enhanced spin
relaxation of injected electrons prior to their thermalization
\cite{Saikin_JPCM06,Mallory_PRB06, Song_CondMat09}. The detrimental
effect of the former can be overcome by increasing the doping
concentration next to the junction. However, the second effect is an
intrinsic property that cannot be engineered for a given zinc-blende bulk
semiconductor. Fig.~\ref{fig:hot_electrons} shows the fraction of
the net injected spin that is left after the energy thermalization
process as a function of the injected energy in a
10$^{16}$~cm$^{-3}$ n-type GaAs \cite{Song_CondMat09}.  The spin
information is largely kept when the energy of injected electrons is
less than 0.1~eV. These results are also in accordance with recent
measured data by Crooker \textit{et al.} \cite{Crooker_PRB09}. The
spin relaxation of these hot electrons is governed by the
Dyakonov-Perel mechanism \cite{Dyakonov_JETP33,
Optical_Orientation}. Due to the moderately reverse biased GaAs/Fe
junction, we have assumed that the injected electrons tunnel into
the $\Gamma$-valley of the conduction band. Thus, the energy
thermalization is governed by emission of long wavelength LO-phonons
\cite{Born_Book}. This scenario is valid if the injected energy of
hot electrons is less than 0.3~eV above the $\Gamma$-point of
conduction band in GaAs. At stronger reverse bias conditions, the
injected electrons can reach the $L$-valley and thus experience
strong inter-valley scattering processes \cite{Yu_Cardona}. In the
case of a 10$^{16}$~cm$^{-3}$ n-type GaAs at room temperature, this
injection energy limit corresponds to a $-$0.4~V reversed biased
GaAs/Fe junction (the Fermi energy is about 0.1~eV below the
conduction band).

We conclude that if the bias voltage across the SC/FM is moderate then one can neglect the spin relaxation processes during the ultra-fast
thermalization of injected electrons to the bottom of the conduction band. This allows one to match the spin polarized tunneling
current densities with the spin polarized
current densities at the edges of the bulk semiconductor region
(Eq.~(\ref{eq:Js})).

\subsection{Intrinsic Capacitance of the Schottky barriers} \label{sec:displacement_currents}

In every moment of time, the applied potential fulfills the Laplace
equation, $ \nabla^2 \phi = \nabla \cdot \mathbf{E} =  0 $ with
boundary conditions given by the charge current densities at the interfaces,
$J_{\alpha}(V)$. In the time-dependent case the charge current density also
includes a displacement current density connected with charging or
discharging the barrier capacitance (changing the width of the
depletion layer),
\begin{equation}
J_{sb} = c_B \frac{\partial V}{\partial t} \,\,.
\label{eq:js_displacement}
\end{equation}
$V$ is the voltage drop across the junction and $c_{B}$ is the
barrier capacitance per unit area whose magnitude is given by the
ratio between the static dielectric constant and the width  of the
Schottky barrier (typically of the order of $10^{-6}$~F/cm$^2$). In
the following simulations, we will include the contribution of this
current as part of the boundary conditions. This current can have a
strong effect on the dynamics in time scales of the order of
$c_B/G$.

Contrary to charge currents, the spin currents are negligibly
affected by the displacement current. This is valid if the change in
the width of the Schottky barrier, $\Delta_d$, is such that $G
l_{sf} / \sigma_0 \gg \Delta_{d}/l_{sf}$. To derive this condition,
we recall that the spin current densities, $\mathfrak{J}^{2D}_{i,\alpha}$,
include terms of the order of $V_T G p_i$. The displacement spin
current density is proportional to $q n_0 \Delta_d\,\partial p_i /
\partial t$. If we are interested in time scales of the order
of or longer than the spin relaxation time ($\,\partial p_i /
\partial t \,< \, p_i / \tau_{sf}\,$) then one can readily derive the above
condition. Neglecting the contribution of the displacement current
is robust even at faster time scales since low voltage signals
change the width of a heavily doped Schottky barrier by only a few
nm  ($\Delta_{d} \ll l_{sf}$). Barriers that are not heavily doped
are of no interest due to the resulting negligible spin
accumulation.

\section{Results and Discussion} \label{sec:RD}

\begin{figure}
\includegraphics[width=7cm]{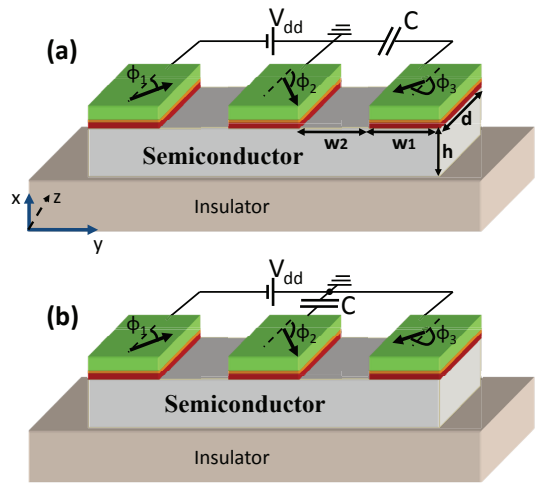}
\caption[setup] {(Color online) The lateral structure we use in the simulations of
Sec.~\ref{sec:RD}. It consists of a semiconductor channel covered by
three ferromagnetic terminals. The geometrical parameters are
$d$=1~$\mu$m, $h$=$w1$=100~nm and $w2$=200~nm. In (a) the capacitor
is connected to the right contact and in (b) to the middle contact.
In both cases, the capacitance is C=4~fF. The magnetic configuration
is denoted by the $\phi_1$, $\phi_2$ and $\phi_3$ angles in the
$yz$-plane (measured from the $+\mathbf{z}$ axis).}
\label{fig:setup}
\end{figure}

In showing the results of the presented theory we focus on
two aspects. First, by using a steady state analysis we show how the spin transport in a semiconductor
channel is affected by the escape current density and by the non-linear J-V relation across the SC/FM junctions. When contrasted with conventional spin transport analyses the results are significantly different . Second, by using a dynamical analysis we study the non-collinear
magnetization configuration effect on the spin accumulation and current
signals. The static and dynamical transport analyses are
performed on a lateral semiconductor channel covered by three
ferromagnetic terminals. Fig.~\ref{fig:setup} shows two bias
settings of the studied structure where in each case two terminals
are biased and a third terminal is connected in series to an
external capacitor (not to be confused with the intrinsic Schottky
capacitance of each SC/FM junction). By perturbing the magnetization
vector of the left or right terminals, the resulting transient charge current across the external capacitor allows
us to study the spin dynamics in the semiconductor channel.

We perform simulations for applied voltage up to 0.3~V with all
possible in-plane magnetization alignments at an interval of $\pi/4$
in each terminal. We recall that the boundary conditions across a
SC/FM junction were derived using the assumption that the spin-$z$
axis is collinear with the majority spin direction in the FM
(sections~\ref{sec:SC/FM} and \ref{sec:doping}). However, in order
to consider all (non-collinear) ferromagnetic terminals and the
semiconductor channel as one system, one should use a single spin
reference coordinate system. Specifically, we transform the general
expressions in Eqs.~(\ref{eq:J_mat_general}) \& (\ref{eq:jz_2D})
into this new `contact-independent' coordinate system. The details
of this transformation as well as the numerical procedure are
explained in Appendix \ref{sec:Numerical Analysis}. We use room
temperature GaAs/Fe material system where the semiconductor
parameters are: $n_0$=10$^{16}$~cm$^{-3}$,
$\nu$=2700~cm$^2$/V$\cdot$s, $\tau_{sf}$=0.2~ns
\cite{Kikkawa_PRL98}, $m_{sc}$=0.067$m_0$. The Fermi wavevectors for
majority and minority electrons in the iron terminals are,
respectively, 1.1~${\AA}^{-1}$ and 0.42~${\AA}^{-1}$ where their mass is of
free electrons \cite{Slonczewski_PRB89}. The doping and static
dielectric constant in the Schottky barrier region are
$n_{sb}$=2$\times$10$^{19}$~cm$^{-3}$ and
$\epsilon_{sc}$=1.16$\times$10$^{-12}$~F/cm, respectively. The
height of the barrier in equilibrium ($V=0$) is $\phi_B$=0.7~eV from
the ferromagnet's Fermi energy. These barrier parameters yield a single localized energy level. In addition,
the combined conductance (of both biased barriers) roughly matches the
semiconductor spin-depth conductance ($\sigma_0/\ell_{sf}\,\sim$4$\times$10$^4$~$\Omega^{-1}$cm$^{-2}$). This
allows us to study cases in which the spin polarization in the
channel is relatively large ($pn_0$ is of the order of $n_0$). Spin
injection experiments, on the other hand, are currently limited to
much smaller polarization values due to the self-compensation issue
of silicon donors in GaAs \cite{Adelmann_JVST05}. This limits the effective interface donor
doping levels to $n_{sb}$$\sim$5$\times$10$^{18}$~cm$^{-3}$
and the resulting conductance at room
temperature to be less than $10^3$~$\Omega^{-1}$cm$^{-2}$ \cite{Hanbicki_APL03}. Breaking
this impasse is a central challenge in the way to realize room
temperature GaAs/Fe spintronics devices.

\subsection{Static Results} \label{sec:Static_Results}

\begin{figure}[htp!]
\includegraphics[width=6.8cm]{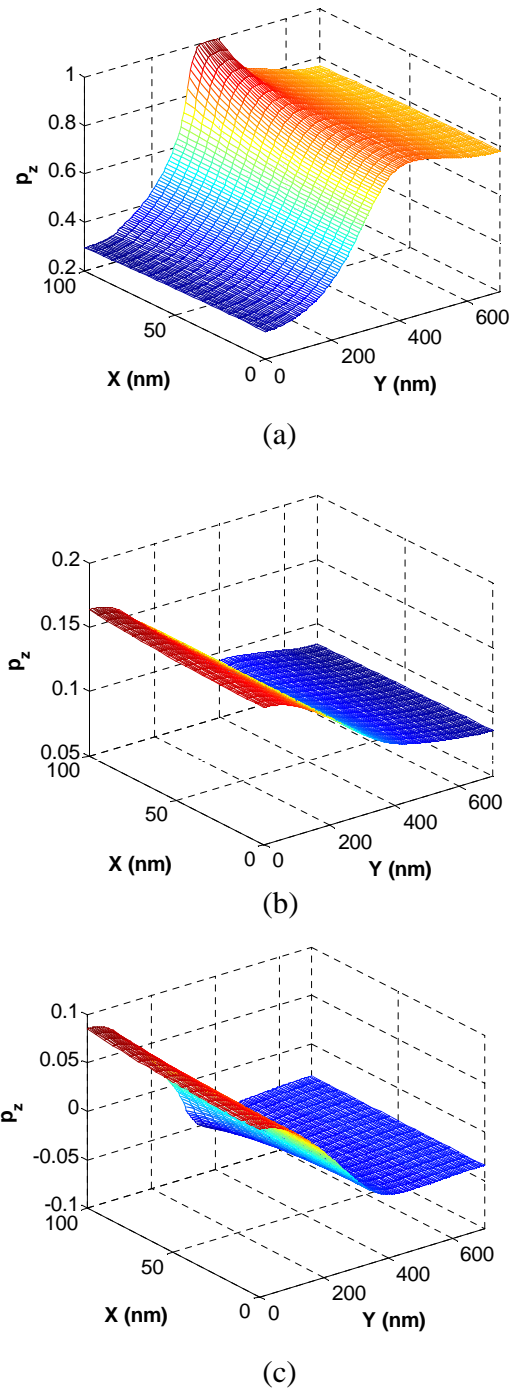}
\caption[static_comparison] {(Color online) Spin polarization in the semiconductor
channel with the setting of Fig.~\ref{fig:setup}(a). The
magnetization directions are all parallel to the $+\mathbf{z}$ axis
and the applied bias is $V_d$=0.3~V. (a) is the result of the full
model, (b) is the result without considering escape current mechanism, and
(c) is the result of the conventional model.}
\label{fig:static_comparison}
\end{figure}

We use a steady state analysis to show the effects of the escape
current density and of the non-linear J-V relation. First, we employ the
setting of Fig.~\ref{fig:setup}(a) in which the right terminal is
outside the path of the steady state charge current. The
magnetization directions of all three contacts are set parallel to
the $+\mathbf{z}$ direction ($\phi_1=\phi_2=\phi_3=0$) so that
$p_z(x,y)$ is the only non-zero spin polarization component. The
applied voltage is $V_{dd}$=0.3~V for which the resulting small voltage drop across each SC/FM junction justifies the use of our boundary conditions. Using
these parameters, Fig.~\ref{fig:static_comparison} shows a direct
comparison of the spin polarization in the semiconductor channel
between (a) the full model, (b) the model without the escape current mechanism
and (c) the conventional model.  Fig.~\ref{fig:static_comparison}(a) shows the result of the full model in which the boundary conditions are given by Eqs.~(\ref{eq:J_mat_general}), (\ref{eq:j0_2D}) \& (\ref{eq:jz_2D}). Fig.~\ref{fig:static_comparison}(b) shows the case when only Eq.~(\ref{eq:J_mat_general}) is employed. Fig.~\ref{fig:static_comparison}(c) shows the results of the widely used
conventional model in which not only the escape current mechanism is ignored but a
linear form of Eq.~(\ref{eq:J_mat_general}) is used. This linear
form is the set of Eqs.~(\ref{eq: interface j0})-(\ref{eq: interface
jsx}) with the replacement of the $e^{-V/V_T}$$-1$ terms by $-V/V_T$.
The conductance and finesses values in the conventional model are extracted
around zero bias. Appendix~\ref{sec:interface current} presents the calculation
of the various reflection coefficient combinations whose integrations provide the conductance and mixing conductance values. In our simulations these values are $F$=0.2,
$G$=2Re$\{G_{\ud}\}$=6Im$\{G_{\ud}\}$=2$\times$10$^4$~$\Omega^{-1}$cm$^{-2}$.

To understand the spin polarization in the semiconductor channel we
consider two scenarios. In case (I), the
transport across a reverse/forward biased junction is dominated by
free/localized electrons. The
net spin polarization produced by a ferromagnetic contact is therefore always
aligned with the majority spin direction. Fig.~\ref{fig:static_comparison}(a) shows the results of this scenario. Case (II) assumes that free
electrons dominate the transport regardless of the bias direction.
In this case, the net spin polarization produced by a reverse/forward biased junction is aligned
with the majority/minority spin direction of the contact. The results of such a scenario are depicted in
Figs.~\ref{fig:static_comparison}(b)~and~(c). In
non-collinear configurations, these rules can be generalized in the
following way. In case (I)/(II), the net spin polarization vector in
the semiconductor channel roughly points in the vector
addition/substraction of the majority spin directions of the biased
junctions.

We first explain the effect of the non-linear J-V relation via a
comparison between the spin polarization in
Figs.~\ref{fig:static_comparison}(b)~and~(c). The most distinct
difference is the shift in values of $p_z(x,y)$. To understand this
shift we recall of the Fermi level positions in the semiconductor
channel and in the ferromagnetic terminal. In a non-degenerate
semiconductor channel the values of the population distribution
matrix are very small ($\hat{f}_{sc}$$<$0.05 in our simulated case).
In very small reverse bias and in any forward bias the population
distribution matrix of the FM, $\hat{f}_{fm}$, is also very small
for electrons that can tunnel to/from the semiconductor. The latter
constraint is removed in higher reverse bias conditions where the
ferromagnetic Fermi level reaches the conduction band edge of the
semiconductor. This is the reason for the free electron asymmetrical current density
at moderate bias conditions as seen from the inset of
Fig.~\ref{fig:well_structure}. The higher conductance of the reverse
biased junction (when considering only free electrons) dictates the
sign of the spin polarization in
Fig.~\ref{fig:static_comparison}(b). On the other hand, in
Fig.~\ref{fig:static_comparison}(c) the spin polarization is nearly
symmetrical about the zero level at the biased part of the channel.
To introduce asymmetry in the conventional model one has to
artificially plug different conductance values for the forward
and reverse biased junctions. However, due to the linear
approximation around zero-bias, the formal derivation of the boundary
conditions in the conventional model results in identical
conductance values.

\begin{figure}[!]
\includegraphics[width=8cm]{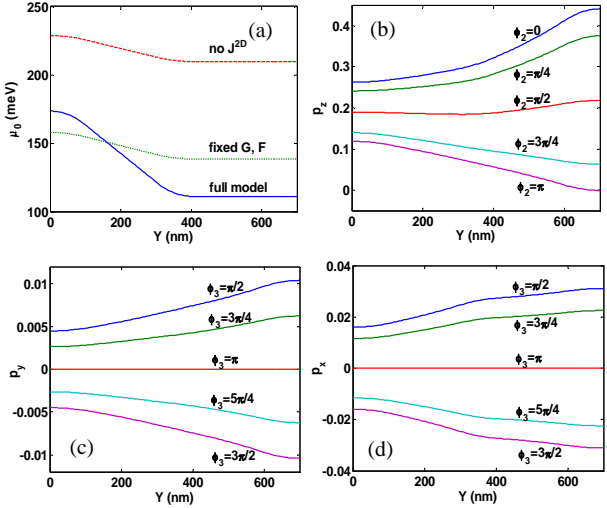}
\caption[1D_static] {(Color online) (a) The averaged electrochemical potential
along the semiconductor channel for the setting of
Fig.~\ref{fig:setup}(a) where $V_{dd}$=0.3~V and
$\phi_1$=$\phi_2$=$\phi_3$=0. (b)-(d) The components of the averaged
spin polarization vector along the semiconductor channel for various
magnetization configurations with the setting of
Fig.~\ref{fig:setup}(b) where $V_{dd}$=0.1~V and $\phi_1$=0. In (b) the middle floating contact is also fixed at $\phi_3$=0. In (c) and (d) the
left biased contact is fixed at $\phi_2$=0.}
\label{fig:1D_static}
\end{figure}

Fig.~\ref{fig:static_comparison}(a) highlights the unique behavior
of the potential well in the interface doping area. We find the
opposite of the conventional result which states that antiparallel
configurations lead to a much larger spin polarization than parallel
configurations \cite{Fert_PRB01, Dery_MCT_PRB07}. In
collinear two-terminal systems the potential well flips the role of
the parallel and antiparallel magnetization configurations (if we
ignore the electrical field effect and assume that the channel
length is smaller than the spin diffusion length.) The potential well significantly changes the shape of the spin accumulation where spins diffuse in opposite directions. This is seen
by the opposite slopes of the spin polarization in the left part of
the channel in Figs.~\ref{fig:static_comparison}(a)~and~(b).
Notably, $p_z$ has a high plateau below the middle forward
biased contact. Since the conductance of the forward biased junction
increases when we incorporate the escape current mechanism, the
relative portion of the voltage drop across the semiconductor
channel increases and as a result the electric field in the channel
increases. The field pushes spins carried by injected electrons to
the $-\mathbf{E}$ direction and thus spins accumulate near the
forward biased side. Studying this high spin-polarization regime
is possible when working with $\mathbf{p}$ rather than with its
direction and with $\mu_{\pm}$ (Eq.~(\ref{eq:mu})). Fig.
\ref{fig:1D_static}(a) shows the $x$-averaged electric potential along
the semiconductor channel, $1/h \int_0^h dx\phi(x,y)$.
From the slope of the curves we can estimate the electrical field
in the left part of the channel to be of the order of 2~kV/cm when incorporating the escape current mechanism and about
four time smaller in the other cases. The resulting drift velocity ($\nu|E|$) is still below the saturation velocity.

To explain the effect of the electric field on the spin currents we
employ the full model with the setting of Fig.~\ref{fig:setup}(b)
and $V_{dd}$=0.1~V. Note that both charge and spin currents can flow
in the semiconductor channel under the floating middle contact.
Figs.~\ref{fig:1D_static}(b)-(d) show the averaged spin components
along the semiconductor channel  ($1/h \int_0^h dx p_i(x,y)$). As
can be seen from Eq.~(\ref{eq:Js}), the signature of the electric
field is evident at fields amplitudes which exceed $V_T/\ell_{sf}$
\cite{Yu_Flatte_long_PRB02} ($>$250~V/cm in non-degenerate n-type
GaAs at room temperature). Fig.~\ref{fig:1D_static}(b) shows the
averaged $p_z$ component for five magnetization directions of the
right biased contact whereas the other two contacts are set parallel
in the $+\mathbf{z}$ direction ($\phi_1=\phi_3=0$). The electric
field opposes the diffusion of spins away from the forward biased
junction and the spin polarization from the reverse biased junction
spreads throughout the channel. This is the reason that even in the
antiparallel configuration ($\phi_2=\pi$), $p_z$ is not much into
negative values beneath the forward biased (right) contact.

Figs.~\ref{fig:1D_static}(c)~and~(d)
show, respectively, the averaged $p_y$ and $p_x$ for $\phi_1=\phi_2=0$. Here
the non-collinearity of the floating middle contact $\phi_3$ (from $\pi/2$ to $3\pi/2$) is the
only drive for the $p_y$ and $p_x$ components. As before, the spin accumulation is pushed by the strong electric field toward the forward biased junction in the right part of the channel. The out-of-plane spin component is a useful probe of spin polarization beneath ferromagnetic contacts. For the chosen setting, we note that $p_y$ which is a direct result of the non-collinear middle contact is smaller than $p_x$. The reason is that $p_x$ is a mixed term that is proportional to cross product of the spin polarization vector and the magnetization direction in the middle contact ($\mathbf{p} \times \hat{\mathbf{m}}_3$). The spin-polarization in the channel is nearly collinear with the $z$-axis because of the magnetization directions of the left and right biased contacts  ($\phi_1$=$\phi_3$=0). Taking, as an example, the case that the magnetization in the floating middle contact is along the $y$-axis ($\phi_3$=$\pi$/2), then the mixed $p_x$ component can be larger than $p_y$ due to the relatively large $p_z$ in the channel. On the other hand, the $p_y$ component is generated by the very small voltage, $V_f$, across the floating contact. This small bias results in a charge current density of magnitude $GV_f$ which due to the external capacitor is contrasted by an equivalent charge current density of magnitude $GV_TF(\mathbf{p}\cdot\hat{\mathbf{m}}_3)$ (nullifying the expression on the right hand side of  Eq.~(\ref{eq: interface j0}) with $V=V_f \rightarrow 0$).

\subsection{Dynamic Results}\label{sec:Dynamic_Results}

\begin{figure}
\includegraphics[width=7cm]{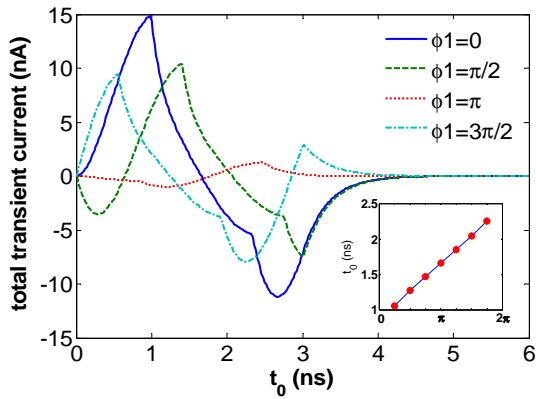}
\caption[3r]
                 {(Color online) Transient currents across the external capacitor due to a single 3~ns clockwise rotation of the magnetization direction of the right terminal. The inset shows the times at which current changes sign from positive to negative as a function of $\phi_1$.
                 } \label{fig:3r}
\end{figure}

We use a dynamical analysis to study how the non-collinearity
affects the current signals. The setting of Fig.~\ref{fig:setup}(a)
is used where the magnetization direction of the middle terminal is
fixed along $+\mathbf{z}$ ($\phi_2$=0). The right terminal is
outside the path of the steady state charge current, and its
magnetization direction is perturbed according to
$\hat{\phi}_3(t)=\cos(2\pi t/\tau_r)\hat{z}+\sin(2\pi
t/\tau_r)\hat{y}$ for $0<t<\tau_r$. The transient current across the
external capacitor is then evaluated for various magnetization
directions of the left terminal ($\phi_1$). Similar dynamical setups
have been suggested for collinear configuration by Cywinski
\textit{et al.} \cite{Cywinski_APL06}. Here, we allow for a more
flexible operation regime (non-collinearity) and we offer a
different physical interpretation of the dynamical results while
emphasizing the robustness of the signals' signature.
Fig.~\ref{fig:3r} shows the transient currents across the external
capacitor for four $\phi_1$ cases. The applied bias is $V_d=$0.1~V,
the external capacitance is $C=4$~fF, and the depth of the system in
$z$ direction is 1~$\mu$m. The magnetization direction of the right
terminal completes a single clockwise rotation in $\tau_r$=3~ns.

The transient currents in Fig.~\ref{fig:3r} are described by
$-(C/q)d \mu_r/dt$ where $\mu_r(t)$ is the potential level in the
right terminal. This current is also the integrated current density
at the right contact. This current density is given by,
\begin{eqnarray}
J_R(t) &=& -\left(G + \frac{n_{2D}\left( 1 - e^{-n_w/n_{2D}}\right)}{2 V_T \tau_{esc,0}} \Theta(V(t)) \right) V(t) \nonumber \\ &\,& \,\,\, - GV_TFp_z(t)
 + c_b\frac{d V(t)}{dt}\,.  \label{eq:transient}
\end{eqnarray}
$V(t)$ denotes the (small) self-adjusted voltage drop across the
right terminal, and $\tau_{esc,0}$ is the escape time at 0 voltage drop. The terms that involve $G$ are from linearizing
Eq.~(\ref{eq: interface j0}) around V=0 and the $c_b$ term is due to
the intrinsic capacitance of the Schottky barrier
(Eq.~(\ref{eq:js_displacement}); we use $c_b$=10$^{-6}$~F/cm$^{2}$
in the simulations). The term that involves the step function,
$\Theta(V(t))$, is due to the escape of localized electrons
(linearizing Eq.~(\ref{eq:j0_2D}) around V=0). $n_w$ denotes the electron density in the potential well and $n_{2D} = qV_Tm_{sc}/\pi \hbar^2$. As discussed at the
end of Sec.~\ref{sec:doping}, our modeling includes transport of
localized electrons only at positive voltages. As a result, the
effective barrier conductance is discontinuous at zero bias. This
discontinuity is the reason for the `cusp' points at times smaller
than 3~ns in Fig.~\ref{fig:3r}. Since we neglect the transport
mechanisms that involve the potential well when $V<$0, this
discontinuity is a model dependent artifact. $p_z(t)$ is the
projected spin polarization vector on $\hat{\phi}_3(t)$. The spin
polarization vector is nearly constant and points in the vector
addition of the majority spin directions of the biased ferromagnetic
terminals (without considering the escape currents it would be the
vector substraction). Thus,
\begin{eqnarray}
p_z(t) \propto  \left\{ {\begin{array}{ccc}
  \cos(\frac{\phi_1}{2}) \cos( \frac{2\pi t}{\tau_r} - \frac{\phi_1}{2}) & , & 0<t<\tau_r=3\,\text{ns} \\
  \cos^2(\frac{\phi_1}{2}) & , & \text{otherwise}
\end{array}}  \label{eq:pz_per} \right.
\end{eqnarray}
For cases that $\phi_1 \neq \{0,\,\pi\}$, the discontinuity in $dp_z(t)/dt$ at t=$\tau_r$ results in an additional `cusp' point
at this time (see Fig.~\ref{fig:3r} at t=3~ns). At times greater than 3~ns,
$J_R$ is governed solely by the dynamics of $V(t)$ towards its
original value prior to the perturbation. At shorter times, $J_R$ is
governed by the counteracting response of $V(t)$ to the perturbing
$p_z(t)$. This response aims at finding a new steady-state condition
and its delay time is dictated by simple circuit analysis (see Fig.~1 in Ref.~\cite{Cywinski_APL06}). If the delay is much longer than the rotation
time then $V(t)$ can be viewed as static. In this case, by
inspection of Eq.~(\ref{eq:transient}) we see that $J_R(t)$ follows
the shape of $p_z(t)$ and its peak reaches an optimal value of $V_T G
F |p|$ (independent of the capacitance). The drawback of using a
long delay is due to the slow dynamics at t$>\tau_r$. On the other
hand, if the delay is very short then $V(t)$ adiabatically follows
$p_z(t)$. However, the resulting peak is now smaller (roughly) by the ratio of
the delay and rotation times.

We use the above analysis to elucidate some of the general features
of the current signals by concentrating on the $\phi_1=0$ case of
Fig.~\ref{fig:3r}. This curve shows two `cusp' points around 1~ns and
2.35~ns. These are the times at which $V(t)$ changes its sign. If the
response time of the system was instantaneous (zero capacitance)
then $V(t)$ would have followed $p_z(t)$ without a delay and these
points would appear in 0.75~ns and 2.25~ns (where $p_z(t)$
changes its sign). The reason for the longer delay in the first point (1~ns - 0.75~ns) than in the second point (2.35~ns - 2.25~ns) is due to the
larger effective barrier conductance in forward bias. The initial
current shape between 0 to 1~ns is due to the initial decrease of
$p_z(t)$ (Eq.~(\ref{eq:pz_per}) with $\phi_1$=0). This change is
counteracted by a 0.25~ns delayed increase of $V(t)$ that tries to
establish a new steady state. In the second branch,
1~ns$<t<$2.35~ns, $V(t)$ is positive in order to counteract the
negative $p_z(t)$. The total transient current begin to decrease due to the turn-on of the escape current process. In the third branch, 2.35~ns$<t<$3~ns, $p_z(t)$ is
positive again and the sudden current drop at 2.35~ns is due to the
stop of the escape current (the $-GV_TFp_z(t)$ component of the
current is counteracted by a weaker and slower response of $V(t)$).
One can repeat this analysis for the other signals in Fig.
\ref{fig:3r}. The difference in their shapes is governed by the
$\phi_1$ phase term of $p_z(t)$. As a result, these signals have an
apparent trend in shifting the time at which the current changes
sign from positive to negative. This crossing time, denoted by
$t_0$, shows a linear dependence in $\phi_1$ as can be seen from the
inset of Fig. \ref{fig:3r}.

\begin{figure}
\includegraphics[width=7cm]{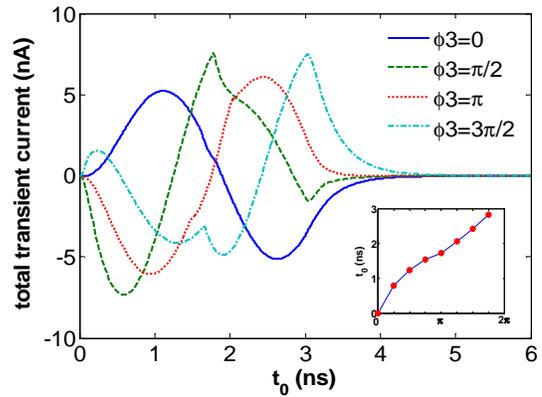}
\caption[1r]
                 {(Color online) Transient currents across the external capacitor due to a single 3~ns clockwise rotation of the magnetization direction of the left terminal.  The inset shows the times at which current changes sign from negative to positive as a function of $\phi_3$.
                 } \label{fig:1r}
\end{figure}

Fig. \ref{fig:1r} shows the current signals for the opposite case in
which the left magnetization is rotated and the right magnetization
is set at various directions (with the same bias setting as before).
We observe similar patterns. In this operational regime, the spin
polarization vector $\mathbf{p}$ changes in time beneath the right
terminal whereas $\phi_3$ is constant. The response time is longer
since the information needs to pass the delay of two (rather than
one) Schottky barriers. For this setting, we denote $t_0$ as the
time at which the signal switches sign from negative to positive. We
observe a similar and nearly linear relation between $\phi_3$ and
$t_0$. However, the slope of the line is about twice as much then
before. This double spacing is best explained with the vector
addition rule we have provided in the static regime. The spin
polarization vector in the semiconductor channel points roughly to
the midway between $\hat{\phi}_2=\hat{z}$ and $\hat{\phi}_1$. Thus,
the rotation speed of $\mathbf{p}$ beneath the floating contact is
about half that of $\phi_1$. In terms of distinguishing different
states, the second setting has a doubled time resolution compared
with the previous case. If we employ $V_d$=0.3~V, then the patterns
of the signal in each setting are very close to the above cases,
while the scale of the signal is about 3 to 5 times larger and the
delay between $p_z(t)$ and $V(t)$ is shorter. Because of its
physical origin, the current signal patterns are very robust and
universal. Discussion about the coercivity and noise concerns can be
found in Ref. \cite{Cywinski_APL06}.

\subsection{Configuration Analysis}\label{sec:Configuration_Analysis}

In this part we explore the maximal number of configurations that can
possibly be stored in a system (as well as how to implement them).
When two biased ferromagnetic contacts have a certain magnetization
alignment, the spin polarization vector results in a unique setting in the semiconductor channel. The vector is more or less
constant throughout the channel if the spin diffusion length is longer than the channel. Thus, this vector labels the particular
magnetization configuration. One way to gain access to this vector
information is to measure the total resistance of the two terminal
system \cite{Brataas_PRL00}. This measurement can tell the relative angle but not the magnetization direction of each contact. In order to extract the information
stored in these magnetization directions completely, we add a third
ferromagnetic contact, floating or semi-floating, on top of the same
semiconductor channel. We use the setting of Fig.~\ref{fig:setup}(a)
where the third magnetization direction provides a reference
direction.

\begin{figure}
\includegraphics[width=8cm]{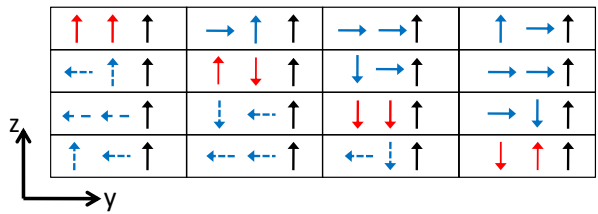}
\caption[configuration_analysis]
                 {(Color online) All possible magnetization configurations with $\phi_3=0$ as a reference direction.  The easy (in plane) axes of the system are in the $y$ and $z$ directions. There are 10 independent configurations in the static case. Each of the `off-diagonal' configurations has a respective symmetric configuration that provides identical voltage drop across the right contact.
                 } \label{fig:configuration_analysis}
\end{figure}

We have shown the physical connection between the measured signal and its
magnetization configuration in the previous part, but we still need a
systematical analysis to quantify the high density storage capacity
in the designated circuit. This analysis is based on symmetry
considerations. We will assume that in each contact there are two mutually perpendicular easy axes,
determined by the semiconductor crystal orientation and by the thickness and shape of the ferromagnetic
contact \cite{Kneedler_PRB97}. Fig. \ref{fig:configuration_analysis} shows all of the possible magnetization
configurations. The right magnetization is
unchanged due to the system rotational
invariance to voltage or current measurement ($\phi_3$ serves
as a reference direction). In principle, there are 10 distinct
voltage output values for these 16 configurations (the output is the voltage difference
between the floating contact and the ground). It is less than 16,
because the voltage of the floating contact is decided by $p_z$ and
$\mu_0$ beneath it and thus the voltage has another invariance when
the magnetization alignment is symmetric about $\mathbf{z}$. The four configurations on the `diagonal' of the $4\times 4$ table in Fig. \ref{fig:configuration_analysis} have a unique output, whereas the `off-diagonal' configurations are symmetric. The degeneracy of the voltage signal of all six `off-diagonal' pairs can be lifted by fixing $\phi_1$ and
$\phi_2$ while turning the $\phi_3$ direction by $\pi/2$ and
repeating the measurement. Each configuration then has a unique
combination of two results. In fact, performing two
static measurements for different $\phi_3$ directions is equivalent
to rotating $\phi_3$ in a given direction and measuring the
transient current dynamically, as showed in the first setting in
Sec. \ref{sec:Dynamic_Results}. In the dynamical method, we can
completely distinguish all 16 states with a single measurement. The
given rotation direction serves to break the symmetry about
$\mathbf{z}$. The dynamic determination scheme is suitable for high frequency operation regime by properly selecting the external capacitor such that the signals are less affected by the noise \cite{Cywinski_APL06, Dery_Nature07}. 

\section{Summary and Outlook} \label{sec:so}
We have presented a detailed model that describes the spin transport in hybrid semiconductor/ferromagnet systems. The derivation of the transport equations and their boundary conditions are then used to model lateral hybrid systems with general non-collinear magnetization configurations. We have corrected the arguments that lead to the use of the quasi-neutrality approximation and explain their momentum and dielectric relaxation times dependence. The spin currents due to tunneling of free electrons across a semiconductor/ferromagnet junction were derived using a rigorous non-linear bias dependence of the tunneling current. The bias voltage limitations of spin injection and extraction were also discussed.

We have introduced an important contribution to the junction tunneling current that is governed by the escape of localized electrons. This process is a result of the usually employed inhomogeneous doping at the vicinity of the semiconductor/ferromagnet interface. The escape current is incorporated to the boundary conditions in forward bias. If the doping inhomogeneity is large then this process becomes the dominant current mechanism and as a result the spin accumulation patterns of parallel and anti-parallel configurations are flipped (compared to the case that only delocalized electrons are considered). We have provided simple rules for estimating the direction and magnitude of the spin-polarization vector in a semiconductor channel that is covered by spin selective and biased ferromagnetic contacts. Our results illustrate the importance of using impedance matched tunneling barriers with the semiconductor spin-depth conductance. This matching condition enables a large spin polarization even if the spin-selective barriers are not ideal. We have introduced a dynamical method that can clearly identify the non-collinear magnetization configuration from a three-terminal lateral structure. The amplitude and pattern of the current signals were explained using a spin dependent circuit analysis that incorporates the capacitive nature of the semiconductor/ferromagnet junction. The presented dynamical method can be used in spintronics devices for storage beyond the binary limit.

This work is supported by DOD/AF/AFOSR FA9550-09-1-0493 and by NSF ECCS-0824075.

\appendix
\section{Barrier Structure Details} \label{sec:interface current}

This appendix elaborates on the model that we use to calculate the
localization energy ($E_i$) and the energy resolved reflection
coefficients ($r_{\uparrow}$ \& $r_{\downarrow}$). These parameters depend on the voltage and are needed for calculating the spin dependent direct and mixing
conductances (Eqs.~(\ref{eq:conductance per
energy}) \& (\ref{eq:mixing conductance per energy})), as well as the
escape times from the potential well (Eq.~(\ref{eq:well_1})).

\begin{figure}
\includegraphics[width=8cm]{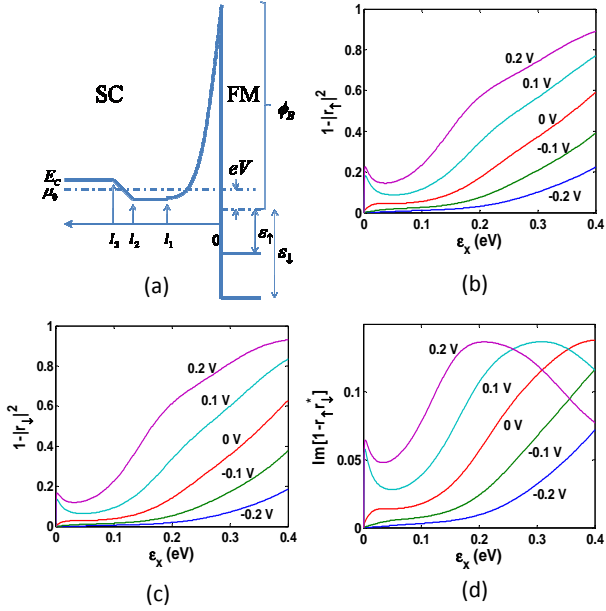}
\caption[reflection]{(Color online) (a) The energy profile of the GaAs/Fe junction
(see text for detailed parameters). (b)-(d), $1-|r_{\ud}|^2$ and
Im$[1-r_{\uparrow}r_{\downarrow}^*]$ versus the longitudinal kinetic
energy for five voltage levels.
                 } \label{fig:reflection}
\end{figure}

Fig.~\ref{fig:reflection}(a) shows the energy profile that we use to
calculate the localization energies and the spin-dependent
reflection coefficients. $V$ is the voltage drop across the
junction. The conduction band energy in the bulk semiconductor is
the reference level ($E_c$=0). The Schottky barrier height is
$\phi_B$=0.7~eV from the Fe Fermi energy  which is
$\varepsilon_{\uparrow}$=4.5~eV ($\varepsilon_{\downarrow}$=0.67~eV)
for majority (minority) electrons from the bottom of the ferromagnetic conduction band \cite{Slonczewski_PRB89}. In the
semiconductor, the doping in the bulk and barrier regions are
$n_0$=10$^{16}$~cm$^{-3}$ and $n_{sb}$=2$\times$10$^{19}$~cm$^{-3}$, respectively. The effective masses are 0.067$m_0$
in GaAs and $m_0$ in Fe. The Schottky barrier is located in the $(0,
\ell_1)$ region where the conduction band is parabolic (obeying the
Poisson equation). The doping inhomogeneity between the bulk and the
interface regions generates a potential well next to the barrier
\cite{Zachau_SSC86}-\cite{Shashkin_Semiconductors02}. We model this
well by a flat potential region from $\ell_1$ to $\ell_2$ and then
by a gradual linear increase to the bulk level from
$\ell_2$ to $\ell_3$. The width of the potential well in its flat
region, $\ell_2-\ell_1$, is governed by the voltage drop across the
junction ($V$). It shrinks/expands with increasing $|V|$ in
reverse/forward bias conditions. This flat region vanishes at a
reverse bias of -0.2~V where $\ell_2=\ell_1=\,$8.5~nm. The depth of
the potential well with respect to the conduction band of the bulk
region is $k_B T \ln(n_{sb}/n_0)$. The gradual region width is
$\ell_3-\ell_2$=4~nm. The geometrical details of this approximated
profile do not substantially change the results of a rigorous
self-consistent Schrodinger-Poisson equation set \cite{Dery_PRL07,
Li_APL09}.

The spin dependence of the reflection coefficients is governed by
the different Fermi velocities in the ferromagnetic region. We
calculate these coefficients by assuming specular transport across
the junction and applying a transfer-matrix method for the above
energy profile (bold line in Fig.~\ref{fig:reflection}(a)). Figs.~\ref{fig:reflection}(b)-(d) show the spin dependent transmission
coefficients and the imaginary part of the mixing term as a function
of the longitudinal energy (kinetic energy along the GaAs/Fe
interface normal). The real part of the mixing term satisfies
2Re$[1-r_{\uparrow}r_{\downarrow}^*]\approx \left(2-|r_{\uparrow}|^2-|r_{\downarrow}|^2\right)$,
in agreement with the conclusion in Ref.~\cite{Brataas_EPJB01}. The
potential well leads to a relatively strong transmission of low
energy free electrons (Ramsauer-Townsend resonance). This is shown
by the peak at low $\epsilon_x$ regime in each of the $V>0$ curves.
The summation over the kinetic energy in the parallel plane smears
this peak in the I-V curve (inset of Fig.~\ref{fig:well_structure}).

To calculate the localization energy we solve the Schrodinger
equation for the semiconductor part of the energy profile (and replace the ferromagnetic part with an infinitely thick barrier whose height is $\phi_B$). This profile
yields a single localized energy level in $V\in$[-0.2~V, 0.2~V] and
its value is essentially linear from -0.096~eV to -0.13~eV. We
then calculate $\tau_{esc}$ (Eq. (\ref{eq:well_1})) and
$\tilde{n}_i$ from $E_i$.

Finally, we present an analytical model for the case of a simple rectangular barrier. This model is then used to extract the direct and mixing conductance values. The reflection coefficient of a rectangular barrier with width $d$, and with barrier height $\phi_B$ from the ferromagnetic potential level ($\mu_0-qV$) is given by
\cite{Ciuti_PRL02},
\begin{eqnarray}
r_{\ud} &=&  - \frac{e^{2\kappa_b d} \gamma_{\ud}^{\ast}\gamma_{sc}
- \gamma_{\ud}\gamma_{sc}^{\ast}} {e^{2\kappa_b d}
\gamma_{\ud}^{\ast}\gamma_{sc}^{\ast} -
\gamma_{\ud}\gamma_{sc}}  \,, \label{eq:r} \\
\kappa_b &=& \sqrt{ \frac{2m_{sc}}{\hbar}(\phi_{B}+\mu_0 - qV
-\varepsilon_{\perp})} \,, \nonumber \\ \gamma_{sc} &=& \kappa_b + i
k_{sc}  \,\,\,\,\,\,,\,\,\,\,\,\, \gamma_{\ud} = \kappa_b + i
\frac{m_{sc}}{m_{fm}}k_{fm,\ud} \,. \nonumber
\end{eqnarray}
The wavevectors $k_{sc}$ \& $k_{fm,\ud}$ are along the normal of the
SC/FM interface. $m_{sc}$ and $m_{fm}$ are, respectively, the electron effective mass in the semiconductor and ferromagnet.
The spin selectivity of the reflection is solely
due to the spin dependent wavevectors in the ferromagnetic contact,
$k_{fm,\ud}$.

Spin injection is possible when two conditions are met,
\begin{eqnarray}
\text{I.}& \,&\,\,\,\,\, e^{\kappa_{0}d} \gg 1 \, , \label{eq:cond1}\\
\text{II.}& \,& \,\,\,\,\, \Delta \equiv \frac{ 2m_{sc}d }{\hbar^2 \kappa_{0}} k_BT < 1  \,. \label{eq:cond2}
\end{eqnarray}
where $\kappa_{0}=\kappa_b(\varepsilon_{\perp}=0)$. In triangular or
parabolic shapes we render the same conditions since the numerical
values of effective $\kappa_{0}$ are somewhat less but of the same
order. The first condition guarantees a resistive rather than an
ohmic contact \cite{Schmidt_PRB00,Rashba_PRB00}. The second
condition guarantees that tunneling is the dominant transport
mechanism across the barrier whereas the thermionic current is
negligible. By increasing the electrons energy, condition II makes
the Boltzmann tail of the population distribution decay faster than
the increase in the transmission coefficient,
$1-|r_{\uparrow(\downarrow)}|^2$. Thus, the main contribution to the
current is from electrons whose energy is at the bottom of the
conduction band. These conditions may also be used to simplify the
calculation of the macroscopic conductances and finesse in
rectangular barriers,
\begin{eqnarray}
G &=&  \frac{q^2}{\hbar \mathcal{A}_0}  \left( \frac{\eta_{\uparrow}}{1 + \eta^2_{\uparrow}} +  \frac{\eta_{\downarrow}}{1 + \eta^2_{\downarrow}}\right)\,   , \label{eq:G} \\
F &=&  \frac{ \left( \eta_{\uparrow}-\eta_{\downarrow} \right)  \left( 1 - \eta_{\uparrow}\eta_{\downarrow} \right)}{ \left(\eta_{\uparrow}+ \eta_{\downarrow}\right)  \left(1 + \eta_{\uparrow}\eta_{\downarrow} \right)} \,, \label{eq:F}\\
G_{\uparrow \downarrow } & = & \frac{G}{2} \left(  1 + i
\frac{\eta_{\uparrow}-\eta_{\downarrow}}{1 +
\eta_{\uparrow}\eta_{\downarrow}} \right)\,,  \label{eq:GMixing}
\end{eqnarray}
where one can see that 2Re$[G_{\uparrow \downarrow }] = G$, and $\mathcal{A}_0$ is defined by,
\begin{eqnarray}
\frac{1}{\mathcal{A}_0} &=&  \frac{\!\!\!\!\sqrt{\kappa_{0}}}{(\pi d)^{3/2}
}e^{\frac{\mu_0}{k_BT} - 2\kappa_{0}d} \frac{1}{\Delta^{-3/2}-1}   ,
\nonumber \\ \eta_{\uparrow(\downarrow)} &=& \frac{ m_{sc}}{m_{fm}}
\frac{k^{\epsilon=0}_{fm,\uparrow(\downarrow)}}{\kappa_0} \,.\nonumber
\end{eqnarray}
In semiconductors with a small electron effective mass and in low bias voltages the finesse
is approximately
$(k_{fm,\uparrow}-k_{fm,\downarrow})/(k_{fm,\uparrow}+k_{fm,\downarrow})$.

\section{Numerical Analysis} \label{sec:Numerical Analysis}

In this appendix, we discuss the techniques we use to solve the
overall set of equations and we provide details about the numerical
scheme. Essentially, we need to solve a system of four equations
(Eq. (\ref{eq:Edivergence0}) \& Eq. (\ref{eq:LinDiffPs})) with
proper boundary and initial conditions. Boundary conditions for both
the static and dynamic cases consist of the continuity of charge
and spin current density. We render the non-approximated form of the
free electrons boundary current densities (Eqs.
(\ref{eq:J_mat_general})-(\ref{eq:mixing conductance per energy})).

We discuss the static case first. When the system is in a steady
state, the charge and spin current densities (Eqs. (\ref{eq:J_0}) \&
(\ref{eq:Js})) have no normal components at the boundaries that are
not in contact with ferromagnetic terminals. In addition, the
external capacitor forces a zero total charge current across the
terminal that is attached to it. For boundaries with biased
ferromagnetic terminals, the normal current density component is
equal to the corresponding current density through the SC/FM
junction. These current densities have been derived in
Sec.~\ref{sec:SC/FM} and Sec.~\ref{sec:doping} under the assumption
that the $z$-axis is collinear with the respective majority spin
direction. However, in order to incorporate the three non-collinear
ferromagnetic terminals and the semiconductor channel in a single
system, we need to work with a contact-independent reference
coordinate system. Specifically, we need to transform the expression
in Eqs.~(\ref{eq:J_mat_general}) and (\ref{eq:jz_2D}) into the new
coordinate system. We rewrite $x$, $y$ and $z$ in these equations as
$\tilde{x}$ , $\tilde{y}$ and $\tilde{z}$ to represent the
contact-dependent coordinate. We reserve $x$, $y$ and $z$ for
coordinates in the contact-independent system. The current density
expressions in these two sets of frames are related by
\begin{eqnarray}
\left( \begin{array}{c} \mathfrak{J}_{x,\alpha} \\\mathfrak{J}_{y,\alpha}\\\mathfrak{J}_{z,\alpha}\end{array} \right)
=\left( \begin{array}{ccc} 1 & 0 &0 \\ 0& \cos\phi & \sin\phi \\ 0 & -\sin\phi & \cos\phi\end{array} \right)
\left( \begin{array}{c} \mathfrak{J}_{\tilde{x},\alpha} \\\mathfrak{J}_{\tilde{y},\alpha}\\\mathfrak{J}_{\tilde{z},\alpha}\end{array} \right)\,,
\end{eqnarray}
Note that the charge current density does not depend on the spin
space coordinate. The components of the spin polarization also need
to be expressed in terms of the components in the
contact-independent system (by the reverse rotation transformation),
\begin{eqnarray}
\left( \begin{array}{c} p_{\tilde{x}} \\p_{\tilde{y}}\\p_{\tilde{z}}\end{array} \right)
=\left( \begin{array}{ccc} 1 & 0 &0 \\ 0& \cos\phi & -\sin\phi \\ 0 & \sin\phi & \cos\phi\end{array} \right)
\left( \begin{array}{c} p_x \\p_y\\p_z\end{array} \right)\,.
\end{eqnarray}

For time dependent simulations, we adopt a similar process except
that we need to consider the displacement current due to capacitance
embedded in the system. This includes the intrinsic capacitance of
the Schottky barrier (Sec.~\ref{sec:displacement_currents}) and the
external capacitor between ground and the semi-floating right
terminal. The initial condition in the dynamical case is the steady
state spin polarization and electrochemical potentials of its
corresponding initial configuration.

We employ a finite difference method to obtain the spin polarization
vector and the electrochemical potential in our multi-terminal
system. The computational grid representing the two-dimensional
semiconductor region has $21\times 141$ nodes with a 5~nm interval.
A system of 4 differential equations is to be solved in this region.
To obtain the desired accuracy, we use two major steps with
iteration methods \cite{Ames, Hoffman}. Step I includes the
evaluation of nonlinear coefficients using the under-relaxation
method. These coefficients are updated periodically. Step II solves
the linearized system of equations using an iterative technique,
where we adopt the successive-over-relaxation method. For time
dependent simulations, we generalize this procedure by using a
Crank-Nicolson implicit method with a time interval of 0.02~ns. The
space and time intervals are adjustable over several orders of
magnitude, and we choose them in consideration of both the result
details and the computation time.


\begin{thebibliography}{99}

\bibitem{Zutic_RMP04} I. {\u Z}uti{\'c}, J. Fabian, and S. Das Sharma, Rev. Mod. Phys. \textbf{76}, 323 (2004).

\bibitem{Johnson_Silsbee_PRB87} M. Johnson and R. H. Silsbee, Phys. Rev. B \textbf{35}, 4959 (1987).
\bibitem{Baibich_PRL88} M. N. Baibich, J. M. Broto, A. Fert, F. Nguyen Van Dau, F. Petroff, P. Etienne, G. Creuzet, A. Friederich, and J. Chazelas, Phys. Rev. Lett. \textbf{61}, 2472 (1988).
\bibitem{Binasch_PRB89} G. Binasch, P. Gr{\"u}nberg, F. Saurenbach, and W. Zinn, Phys. Rev. B \textbf{39}, R4828 (1989).
\bibitem{Moodera_PRL95} J. S. Moodera, L. R. Kinder, T. M. Wong, and R. Meservey, Phys. Rev. Lett. \textbf{74}, 3273 (1995).
\bibitem{Prinz_Science98} G. A. Prinz, Science \textbf{282}, 1660 (1998).

\bibitem{Zhu_PRL01} H. J. Zhu, M. Ramsteiner, H. Kostial, M. Wassermeier, H.-P. Schonherr, and K. H. Ploog, Phys. Rev. Lett. \textbf{87}, 016601 (2001).
\bibitem{Hanbicki_APL02} A. T. Hanbicki, B. T. Jonker, G. Itskos, G. Kioseoglou, and A. Petrou, Appl. Phys. Lett. \textbf{80}, 1240 (2002).
\bibitem{Jonker_IEEE03} B. T. Jonker, Proceedings of the IEEE \textbf{91}, 727 (2003).
\bibitem{Hanbicki_APL03} A. T. Hanbicki, O. M. J. van 't Erve, R. Magno, G. Kioseoglou, C. H. Li, B. T. Jonker, G. Itskos, R. Mallory, M. Yasar, and A. Petrou, Appl. Phys. Lett. \textbf{82}, 4092 (2003).
\bibitem{Adelmann_JVST05} C. Adelmann, J. Q. Xie, C. J. Palmstr{\o}m, J. Strand, X. Lou, J. Wang, and P. A. Crowell, J. Vac. Sci. Technol. B \textbf{23}, 1747 (2005). 
\bibitem{Jiang_PRL05}  X. Jiang, R. Wang, R. M. Shelby, R. M. Macfarlane, S. R. Bank, J. S. Harris, and S. S. P. Parkin, Phys. Rev. Lett. \textbf{94}, 056601 (2005).
\bibitem{Appelbaum_Nature07} I. Appelbaum, B. Q. Huang, and D. J. Monsma, Nature \textbf{447}, 295 (2007). 
\bibitem{Jonker_NaturePhys} B. T. Jonker, G. Kioseoglou, A. T. Hanbicki, C. H. Li, and P. E. Thompson, Nature Physics \textbf{3}, 542 (2007).
\bibitem{Min_NatureMaterial06} B.-C. Min, K. Motohashi, C. Lodder, and R. Jansen, Nature Materials \textbf{5}, 817 (2006).
\bibitem{Hueso_Nature07} L. E. Hueso, J. M. Pruneda, V. Ferrari, G. Burnell, J. P. Vald\'{e}s-Herrera, B. D. Simons, P. B. Littlewood, E. Artacho, A. Fert, and N. D. Mathur, Nature \textbf{445}, 410 (2007).
\bibitem{Appelbaum_PRL07} B. Q. Huang, D. J. Monsma, and I. Appelbaum, Phys. Rev. Lett. \textbf{99}, 177209 (2007).


\bibitem{Butler_JAP97} W. H. Butler, X.-G. Zhang, X. Wang, J. van Ek, and J. M. MacLaren, J. Appl. Phys. \textbf{81}, 5518 (1997).
\bibitem{Wunnicke_PRB02} Ph. Mavropoulos, O. Wunnicke and P. H. Dederichs, Phys. Rev. B  \textbf{66}, 024416 (2002).
\bibitem{Mavropoulos_PRB02} O. Wunnicke, Ph. Mavropoulos, R. Zeller, P. H. Dederichs and D. Grundler, Phys. Rev. B  \textbf{65}, 241306(R) (2002).
\bibitem{Zhao_PRB02} Y. J. Zhao, W. T. Geng, A. J. Freeman, and B. Delley, Phys. Rev. B \textbf{65}, 113202 (2002).
\bibitem{Zwierzychi_PRB03} M. Zwierzycki, K. Xia, P. J. Kelly, G. E. W. Bauer and I. Turek, Phys. Rev. B  \textbf{67}, 092401 (2003).
\bibitem{Zega_PRL06} T. J. Zega, A. T. Hanbicki, S. C. Erwin, I. {\u Z}uti{\'c}, G. Kioseoglou, C. H. Li, B. T. Jonker, and R. M. Stroud, Phys. Rev. Lett. \textbf{96}, 196101 (2006).
\bibitem{Chantis_PRL07} A. N. Chantis, K. D. Belashchenko, D. L. Smith, E. Y. Tsymbal, M. van Schilfgaarde, and R. C. Albers, Phys. Rev. Lett. \textbf{99}, 196603 (2007).
\bibitem{Honda_PRB08} S. Honda, H. Itoh, J. Inoue, H. Kurebayashi, T. Trypiniotis, C. H. W. Barnes, A. Hirohata, and J. A. C. Bland, Phys. Rev. B \textbf{78}, 245316 (2008).


\bibitem{Schmidt_PRB00} G. Schmidt, D. Ferrand, L. W. Molenkamp, A. T. Filip, and B. J. van Wees Phys. Rev. B {\bf 62}, R4790 (2000).
\bibitem{Rashba_PRB00} E. I. Rashba, Phys. Rev. B \textbf{62}, R16267 (2000).
\bibitem{Fert_PRB01} A. Fert and H. Jaffr{\`es}, Phys. Rev. B  \textbf{64}, 184420 (2001).
\bibitem{Smith_PRB01} D. L. Smith and R. N. Silver, Phys. Rev. B \textbf{64}, 045323 (2001).
\bibitem{Yu_Flatte_long_PRB02} Z. G. Yu and M. E. Flatt{\'e}, Phys. Rev. B {\bf 66}, 235302 (2002).
\bibitem{Albrecht_PRB03} J. D. Albrecht and D. L. Smith, Phys. Rev. B {\bf 68}, 035340 (2003).
\bibitem{Dery_lateral_PRB06} H. Dery, {\L} Cywi{\'n}ski, and L. J. Sham, Phys. Rev. B {\bf 73}, 041306(R) (2006).
\bibitem{Rashba APL02} E. I. Rashba, Appl. Phys. Lett. \textbf{80}, 2329 (2002).
\bibitem{Cywinski_APL06}  {\L} Cywi{\'n}ski, H. Dery, and L. J. Sham, Appl. Phys. Lett. {\bf 89}, 042105 (2006).
\bibitem{Dery_Nature07} H. Dery, P. Dalal, {\L} Cywi{\'n}ski and L. J. Sham, Nature  \textbf{447}, 573 (2007).


\bibitem{Shockley_BSTJ49} W. Shockley, G. L. Pearson, and J. R. Haynes, Bell System Tech. J. \textbf{28}, 344 (1949).
\bibitem{Herring_BSTJ49} C. Herring, Bell System Tech. J. \textbf{28}, 401 (1949).
\bibitem{Bardeen_BSTJ49} J. Bardeen, Bell System Tech. J. \textbf{28}, 428 (1949).
\bibitem{Brooks_TR53} H. Brooks, Technical report No. \textbf{181}, Cruft Laboratory, Harvard University (Cambridge, Massachusetts, 1953).
\bibitem{Roosbroeck_PR61} W. Van Roosbroeck, Bell System Tech. J. \textbf{29}, 560 (1950); Phys. Rev. \textbf{123}, 474 (1961).



\bibitem{Johnson_Science93} M. Johnson, Science \textbf{260}, 320 (1993).
\bibitem{Jedema_Nature01} F. J. Jedema, A. T. Filip, and B. J. van Wees, Nature {\bf 410}, 345 (2001).
\bibitem{Crooker_Science05} S. A. Crooker, M. Furis, X. Lou, C. Adelmann, D. L. Smith, C. J. Palmstr{\o}m, and P. A. Crowell, Science \textbf{309}, 2191 (2005).
\bibitem{Lou_PRL06} X. Lou, C. Adelmann, M. Furis, S. A. Crooker, C. J. Palmstr{\o}m, and P. A. Crowell, Phys. Rev. Lett. {\bf 96}, 176603 (2006).
\bibitem{Lou_NaturePhys07} X. Lou, C. Adelmann, S. A. Crooker, E. S. Garlid, J. Zhang, K. S. Madhukar Reddy, S. D. Flexner, C. J. Palmstr{\o}m, and P. A. Crowell, Nature Phys. \textbf{3}, 197 (2007).
\bibitem{Ji_APL06} Y. Ji, A. Hoffman, J. E. Pearson, and S. D. Bader, Appl. Phys. Lett. \textbf{88}, 052509 (2006).
\bibitem{Saha_APL07} D. Saha, M. Holub, and P. Bhattacharya, Appl. Phys. Lett. \textbf{91}, 072513 (2007).
\bibitem{Erve_APL07} M. J. van 't Erve, A. T. Hanbicki, M. Holub, C. H. Li, C. Awo-Affouda, P. E. Thompson, and B. T. Jonker, Appl. Phys. Lett. \textbf{91}, 212109 (2007).
\bibitem{Crooker_PRB09} S. A. Crooker, E. S. Garlid, A. N. Chantis, D. L. Smith, K. S. M. Reddy, Q. O. Hu, T. Kondo, C. J. Palmstr{\o}m, and P. A. Crowell, Phys. Rev. B \textbf{80}, 041305(R) (2009).


\bibitem{Tunneling_Phenomena}  {\it Tunneling Phenomena in Solids}, ed. E.~Burstein and S.~Lundqvist (Plenum Press, New York, 1969).
\bibitem{Sze} S. M. Sze, {\it Physics of Semiconductor Devices} (John Wiley, New York, 1981).

\bibitem{Zhang_PRB02} S. Zhang and P. M. Levy, Phys. Rev. B \textbf{65}, 052409 (2002).
\bibitem{Zhang_PRB05} J. Zhang and P. M. Levy, Phys. Rev. B \textbf{71}, 184417 (2005).
\bibitem{Zhu_PRB08} Y. H. Zhu, B. Hillebrands, and H. C. Schneider, Phys. Rev. B  \textbf{78}, 054429 (2008).

\bibitem{Slonczewski_PRB89} J. C. Slonczewski, Phys. Rev. B \textbf{39}, 6995 (1989).
\bibitem{Ustinov_JPCM95} V. Ustinov and E. Kravtso, J. Phys. Condens. Matter \textbf{7}, 3471 (1995).
\bibitem{Camblong_PRB51} H. E. Camblong, P. M. Levy and S. Zhang, Phys. Rev. B \textbf{51}, 16052 (1995).
\bibitem{Valet_PRB93}  T. Valet and A. Fert, Phys. Rev. B {\bf 48}, 7099 (1993).
\bibitem{Brataas_PRL00} A. Brataas, Y. V. Nazarov, and G. E. W. Bauer, Phys. Rev. Lett.  \textbf{84}, 2481 (2000).
\bibitem{Hernando_PRB00} D. H. Hernando, Y. V. Nazarov, A. Brataas, and G. E. W. Bauer, Phys. Rev. B  \textbf{62}, 5700 (2000).
\bibitem{Brataas_EPJB01} A. Brataas, Y. V. Nazarov, and G. E. W. Bauer, Eur. Phys. J. B  \textbf{22}, 99 (2001).
\bibitem{Xu_Nanotechnology08} Y. Xu, K. Xia, and Z. Ma,  Nanotechnology \textbf{19}, 235404 (2008).


\bibitem{Saikin_JPCM04} S. Saikin, J. Phys. Condens. Matter  \textbf{16}, 5071 (2004).
%
\bibitem{Yang_neutrality} Y. Song, J. Galkowski, and H. Dery (to be submitted for publication).


\bibitem{Ciuti_PRL02} C. Ciuti, J. P. McGuire, and L. J. Sham, Phys. Rev. Lett.  \textbf{89}, 156601 (2002).
\bibitem{Kawakami_Science01} R. K. Kawakami, Y. Kato, M. Hanson, I. Malajovich, J. M. Stephens, E. Johnston-Halperin, G. Salis, A. C. Gossard, D. D. Awschalom, Science \textbf{294}, 131 (2001).
\bibitem{Epstein_PRB02} R. J. Epstein, I. Malajovich, R. K. Kawakami, Y. Chye, M. Hanson, P. M. Petroff, A. C. Gossard, and D. D. Awschalom, Phys. Rev. B \textbf{65}, 121202(R) (2002).
\bibitem{Li_PRL08} Y. Li, Y. Chye, Y. F. Chiang, K. Pi, W. H. Wang, J. M. Stephens, S. Mack, D. D. Awschalom, and R. K. Kawakami, Phys. Rev. Lett. \textbf{100}, 237205 (2008).

\bibitem{Osipov_PRB04} V. V. Osipov and A. M. Bratkovsky, Phys. Rev. B  \textbf{70}, 205312 (2004).
\bibitem{Dery_PRL07} H. Dery and L. J. Sham, Phys. Rev. Lett.  \textbf{98}, 046602 (2007).
\bibitem{Li_APL09} P. Li and H. Dery, Appl. Phys. Lett. \textbf{94}, 192108 (2009).
\bibitem{Smith_PRB08} D. L. Smith and P. P. Ruden, Phys. Rev. B \textbf{78}, 125202 (2008).

\bibitem{Zachau_SSC86} M. Zachau, F. Koch, K. Ploog, P. Roentgen, and H. Beneking,  {\it et al.},  Solid State Commun. \textbf{59}, 591 (1986).
\bibitem{Geraldo_JAP93} J. M. Geraldo, W. N. Podrigues, G. Medeiros-Ribeiro, and A. G. de Oliveira, J. Appl. Phys. \textbf{73}, 820 (1993).
\bibitem{Shashkin_Semiconductors02}  V. I. Shashkin, A. V. Murel, V. M. Daniltsev, and O. I. Khrykin, Semiconductors \textbf{36}, 505 (2002).

\bibitem{Deveaud_APL88} B. Deveaud, J. Shah, T. C. Damen, and W. T. Tsang, Appl. Phys. Lett. \textbf{52}, 1886 (1988).
\bibitem{Kuhn_chapter} T. Kuhn, in {\it Theory of Transport Properties of Semiconductor NanoStructures}, edited by E. Sch\"{o}ll (Chapman \& Hall, London, 1998), pp. 173–214.
\bibitem{Dery_PRB03} H. Dery, B. Tromborg and G. Eisenstein, Phys. Rev. B \textbf{67}, 245308 (2003).

\bibitem{Malinowski_PRB00} A. Malinowski, R. S. Britton, T. Grevatt, R. T. Harley, D. A. Ritchie, and M. Y. Simmons, Phys. Rev. B \textbf{62}, 13034 (2000).

\bibitem{Weng_PRB04} M. Q. Weng, M. W. Wu, and L. Jiang, Phys. Rev. B {\bf 69}, 245320 (2004). 

\bibitem{Saikin_JPCM06} S. Saikin, M. Shen, and M. C. Cheng, J. Phys. Condens. Matter  \textbf{18}, 1535 (2006).
\bibitem{Mallory_PRB06} R. Mallory, M. Yasar, G. Itskos, A. Petrou, G. Kioseoglou, A. T. Hanbicki, C. H. Li, O. M. J. van't Erve, B. T. Jonker, M. Shen and S. Saikin, Phys. Rev. B \textbf{73}, 115308  (2006).
\bibitem{Song_CondMat09} Y. Song and H. Dery, cond-mat/0909.3124

\bibitem{Dyakonov_JETP33} M. I. Dyakonov and V. I. Perel, Sov. Phys. JETP \textbf{33}, 1053 (1971); Sov. Phys. Solid State \textbf{13}, 3023 (1972).
\bibitem{Optical_Orientation} G. E. Pikus and A. N. Titkov, in \textit{Optical Orientation}, edited by F. Meier and B. P. Zakharchenya, Vol. 8, 73-131 (Nort-Holland, New York, 1984).
\bibitem{Born_Book} M. Born and K. Huang, {\it Dynamical Theory of Crystal Latticess} (Oxford University Press, Oxsford, England, 1988).
\bibitem{Yu_Cardona} P. Y. Yu and M. Cardona, {\it Fundamentals of Semiconductors} (Springer, Berlin, 1996).


\bibitem{Kikkawa_PRL98} J. M. Kikkawa and D. D. Awschalom, Phys. Rev. Lett. \textbf{80}, 4313 (1998). 
\bibitem{Dery_MCT_PRB07} H. Dery, L. Cywinski, and L.J. Sham, Phys. Rev. B \textbf{73}, 161307(R) (2006). 

\bibitem{Kneedler_PRB97} E. M. Kneedler, B. T. Jonker, P. M. Thibado, R. J. Wagner, B. V. Shanabrook, and L. J. Whitman, Phys. Rev. B \textbf{56}, 8163 (1997).

\bibitem{Ames} W. F. Ames, {\it Numerical Methods for Partial Differential Equations}, 2nd ed. (Academic, New York, 1977).
\bibitem{Hoffman} J. D. Hoffman, {\it Numerical Methods for Engineers and Scientists}, 2nd ed. (CRC, New York, 2001).








\end{thebibliography}

\end{document}